# Stochastic Modeling of 3-D Compositional Distribution in the Crust with Bayesian Inference and Application to Geoneutrino Observation in Japan

**N. Takeuchi[1], K. Ueki[1,2], T. Iizuka[3], J. Nagao[3], A. Tanaka[4] , S. Enomoto[5,6], Y. Shirahata[7], H. Watanabe[7], M. Yamano[1] and H. K. M. Tanaka[1]**

[1]Earthquake Research Institute, University of Tokyo, Bunkyo-ku, Tokyo 113-0032, Japan.

[2] Department of Solid Earth Geochemistry, Japan Agency for Marine-Earth Science and Technology, Yokosuka 237-0061, Japan.

[3] Department of Earth and Planetary Science, University of Tokyo, Bunkyo-ku, Tokyo 113-0033, Japan.

[4] Geological Survey of Japan, National Institute of Advanced Industrial Science and Technology, Tsukuba 305-8567, Japan

[5] Kavli Institute for the Physics and Mathematics of the Universe, University of Tokyo, Kashiwa 277-8583, Japan.

[6] Department of Physics, University of Washington, Seattle, WA 98195, United States.

[7] Research Center for Neutrino Science, Tohoku University, Sendai 980-8578, Japan.

Corresponding author: Sanshiro Enomoto (sanshiro@uw.edu)

**Highlights**

- Objective method for crustal modeling based on Bayesian inference is developed and applied to Japan arc.
- Biases in conventional geochemical methods are investigated and a new bias-free rock composition model is presented.
- Geo-neutrino flux from the Japan crust is calculated with fully probabilistic uncertainty estimation.

## Abstract

Geoneutrino observations, first achieved by KamLAND in 2005 and followed by Borexino in 2010, have accumulated statistics and improved sensitivity for more than ten years. The uncertainty of the geoneutrino flux at the surface is now reduced to a level small enough to set useful constraints on U and Th abundances in the bulk silicate earth (BSE). However, in order to make inferences on earth's compositional model, the contributions from the local crust need to be understood within a similar uncertainty. Here we develop a new method to construct a stochastic crustal composition model utilizing Bayesian inference. While the methodology has general applicability, it incorporates all the local uniqueness in its probabilistic framework. Unlike common approaches for this type of problem, our method does not depend on crustal segmentation into upper, (middle) and lower, whose classification and boundaries are not always well defined. We also develop a new modeling method to infer rock composition distributions that conserve mass balance and therefore do not bias the results. Combined with a new vast collection of geochemical data for rock samples in the Japan arc, we apply this method to geoneutrino observation at Kamioka, Japan. Currently a difficulty remains in the handling of correlations in the flux integration; we conservatively assume maximum correlation, which leads to large flux estimation errors of 60~70%. Despite the large errors, this is the first local crustal model for geoneutrino flux prediction with probabilistic error estimation in a reproducible way.

## 1 Introduction

Recent developments on detection of low energy antineutrinos with large scintillation detectors, such as the KamLAND detector (Eguchi et al., 2003) in Kamioka, Japan, and the Borexino detector (Alimonti et al., 2009) in Gran Sasso, Italy, have demonstrated their ability to detect antineutrinos from radioactive decays inside the earth (geoneutrinos) (Araki et al., 2005; Bellini et al., 2010). Among beta-decaying isotopes abundant in the earth, antineutrinos from $^{238}$U and $^{232}$Th decay chain elements are detectable by those detectors (Table 1 in (Enomoto, 2006)). Although geoneutrinos are expected to bring unique knowledge about energetics and composition of the deep interior of the earth (e.g., Krauss et al., 1984, Kobayashi & Fukao, 1991; Raghavan et al., 1998; Mantovani et al., 2004; Enomoto et al., 2007; Šrámek et al., 2013), extraction of such information from geoneutrino observations at the surface requires detailed understanding of local crustal contributions.

As of 2017, the latest KamLAND geoneutrino observation (available at the KamLAND web site http://www.awa.tohoku.ac.jp/kamland/?p=2664, accessed August 2017) measures the geoneutrino flux with an uncertainty of 18% (U/Th mass ratio fixed). Compared to the quoted "subjective" uncertainty of the Bulk Silicate Earth (BSE) model composition (McDonough & Sun, 1995), 20% (U) and 15% (Th), the KamLAND measurement demonstrates that geoneutrino observations are now able to set useful constraints on earth models. As shown in Enomoto et al. (2007), geoneutrinos from the Japan crust contributes approximately half of the total flux observed at Kamioka. Therefore, in order to use the geoneutrino observation to make inferences on global Earth science, the local crustal contribution needs to be understood with better accuracy than the flux observation errors; or at least, the local contribution must be modeled with a quantitative estimation of its uncertainties.

To estimate the local crustal contribution, one needs to model the distributions of U and Th in the entire local crust with quantitative error estimation. So far, several attempts to this end have been made. For the Kamioka vicinity, two studies were made using a surface geological map (Geological Survey of Japan, 1995) and a large number of rock samples associated with the geological map (Togashi et al., 2000). The first study by Fiorentini et al. (2005) used the surface map to estimate the local crustal contribution to be added to their global model flux. The second study, Enomoto (2005), used the surface map to evaluate the model errors originating in the local crustal uniqueness and heterogeneity to be added to their global model uncertainties. Both studies assumed that the surface exposed geology extends vertically to a certain depth. Both used the seismic topography models of Moho and Conrad by Zhao et al. (1992) for segmentation of the upper and lower crusts, and both adopted a homogenous lower crust model. For the Gran Sasso vicinity, where the Borexino detector is located, Coltorti et al. (2011) estimated the local crustal contribution with segmented seismological profiles of the crust combined with surface exposed samples. For the compositions of the local upper and lower crusts, the felsic/mafic crust volume ratio was inferred from seismic velocity, with a binary grouping of all rock types into felsic or mafic. Their local model does not include the middle crust although their global model does.

A similar felsic/mafic binary model was later used by Huang et al. (2013) for global crustal modeling. They modeled the middle crust as a linear mixture of felsic and mafic amphibolite facies rocks, and the lower crust as a linear mixture of felsic and mafic granulite facies rocks, then determined the felsic/mafic crust ratios using Vp in the CRUST2.0 (Laske et al., 2001) seismic velocity model. This method was not adopted to the upper crust in Huang et al. (2013), although two of the five authors are from Coltorti et al. (2011). Instead, the global upper crustal composition by Rudnick and Gao (2003) was used assuming a globally homogeneous upper crust. Errors of crustal thickness were estimated by comparing several seismic crustal models.

As for the upper crust, Huang et al. (2014) estimated its U and Th distributions for the Sudbury vicinity, where a new large-volume scintillation detector, SNO+ (Andringa et al., 2016), is being built. The flux from the local upper crust was calculated by assigning local surface rock sample compositions to their 3-D geological map constructed by combining the surface geological map and several seismological and geological profiles.

All of the local and global models constructed so far estimate uncertainties by comparing different input models (seismic crustal boundaries, compositional models, etc.), or by picking an arbitrary (large) error, for some parts of the calculations if not all. Needless to say, there is no guarantee that constructing a probability density function (PDF) by collecting published values will constitute the correct probability distribution. Also the choice of inputs is often arbitrary. Some 3-D models were constructed by interpolating or extrapolating 2-D surface geological maps, 2-D seismic cross sections, and/or 1-D bore-hole data into 3-D distributions, without estimating uncertainties associated with such treatments. The boundaries of crustal layers, the depths of layers, or even the very existence of boundaries (especially for the middle crust) are highly uncertain, as can be seen in variations of crustal boundary models and in mixtures of 2-layer and 3-layer models.

For rock composition models, the log-normal distribution has been widely used in geochemistry including geoneutrino modeling studies (Huang et al., 2013; Huang et al., 2014; Šrámek et al., 2016). Adopting a log-normal distribution for rock compositions was suggested by Ahrens (1954), based on their observation that igneous rocks commonly show skewed compositional distributions that do not fit to a normal distribution. Chayes (1954) immediately responded to this, arguing that the log-normal model is not a unique deduction from the skewness and hence the logic is inadequate. Furthermore, it can be shown that if a log-normal PDF is fit to a sample distribution with the method of maximum likelihood (or, equivalently, maximum a posteriori (MAP) with a flat prior), the mean value (expected value of the fit PDF) will be biased, resulting in a bias in the final flux integration. In log-normal models, maximum likelihood estimation of the geometric mean is unbiased, but it is in general different from the arithmetic mean.

Another source of bias is the reliance on median, which is widely used as a representation of rock compositions or bulk reservoir compositions. For positively skewed distributions, median is smaller than the mean, and if median values are used in integration or in any mass-balance calculations, the result will have neither the correct mean nor the correct median, as median is not an additive quantity. Further, if "filtering around the median" is applied to samples to remove "atypical" samples in distribution tails, which is also a common treatment in geochemistry, the mean of the filtered samples depends on the filter cut-off and hence the resultant mean becomes arbitrary. This can be avoided if distribution shape information is properly used, though such treatment is not common.

In this work, we develop a method of 3-D crustal modeling with uncertainty estimation that is truly probabilistic and reproducible, as opposed to "uncertainties by comparison" or "subjective / arbitrary estimation". In our method, we consistently use explicit PDFs for all relevant quantities. We utilize Bayesian inference techniques to model the 3-D lithology map by combining seismological data as "observation" with a prior model constructed from local exposure. By using seismological tomography, we avoid the difficulty of dealing with the upper / middle / lower crust classification and boundary definition. For rock composition, we adopt a gamma distribution model, which does not bias the mean value estimation (unlike the log-normal model) and fits consistently well to both highly-skewed and close-to-normal distributions (for which neither log-normal nor normal distributions apply). Although the gamma model is still not a unique deduction as by Chayes (1954), it is at least relatively a better choice than log-normal or normal, and we see interesting characteristics of the gamma model that suggest some underlying geological processes. We also develop a method to properly treat samples below detection limits. The framework developed here is populated by a new vast collection of local rock samples, and is applied to geoneutrino observation at Kamioka, Japan.

## 2 Overview of the method

Fig. 1 shows a flow chart of the neutrino flux calculation in this study. The flux calculation is conducted in three steps: inference of the lithology map, inference of U and Th concentrations for each rock type, and inference of geoneutrino flux at the surface. The first, the second, and the last step are discussed in Sections 3, 4, and 5, respectively.

In the first step, we first assume a prior probability map of lithology $P^{(\boldsymbol{x})}(i)$, the probability that rock at location $\boldsymbol{x}$ is type $i$. This prior probability allows us to take local lithological features into account. In this study, to apply our method to calculate the flux at Kamioka, Japan, we use the lithology of the Hidaka metamorphic belt in Hokkaido, northern Japan, as the prior. Resulting from the obduction of the Kuril arc onto the Northeast Japan arc, the Hidaka metamorphic belt provides an arc sequence exposure that encompasses shallow crust to the uppermost mantle (e.g., Kimura, 1986; Komatsu et al., 1989; Osanai et al., 1991). The Kuril arc constitutes only a small portion of the Japan arc system, yet it has similarities with the main Japan arcs in tectonic histories where they initially evolved at the continental margin through subduction-related magmatism and accretion, followed by back-arc basin formation (Maruyama et al., 1997; Taira, 2001). Thus, the crustal section of the Hidaka metamorphic belt is expected to represent a typical example of crustal lithology in the Japan arc system. We use the proportion of rocks constituting the Hidaka crustal section as the prior probability model for every location $\boldsymbol{x}$; here we remove the information on depth distribution of rocks from the prior and leave the seismic data to resolve it.

We then update the probability using Bayes theorem. We refer to a seismic tomography model to get the P velocity data at $\boldsymbol{x}$, $v^{obs}$, and compute the posterior probability map $P^{(\boldsymbol{x})}(i|v^{obs})$, which constrains $i$ to those rock types for which the P velocity is likely to be around $v^{obs(\boldsymbol{x})}$. Note that, in contrast to previous studies, we did not use seismological models with somewhat *ad-hoc* segmentations of, for example, upper, middle and lower crusts. The use of the tomography model obtained in a reproducible manner allows us to evaluate the errors of the seismological data straightforwardly. To compute the posterior probability map $P^{(\boldsymbol{x})}(i|v^{obs})$, we develop an appropriate Bayesian velocity-lithology translator. We refer to laboratory experiments by Christensen and Mooney (1995) for P velocities of various rock types in the crust.

In the second step, we construct U and Th concentration PDFs for each rock type, $f_{\mathrm{U}}^{(i)}(a)$ and $f_{\mathrm{Th}}^{(i)}(a)$, where $a$ denotes the concentration. In previous studies, Togashi et al. (2000) showed slight depletion in incompatible elements (including U and Th) in the average Japan upper crust compared to typical continental upper crusts. Hacker et al. (2015) pointed out that arc lower crust is depleted in incompatible elements compared to typical continental lower crusts. Considering these, we collect U and Th concentration data of Japanese rock samples to incorporate possible local compositional deviations, rather than just using the compositions of typical continental crusts. Based on the collected data, we construct a composition PDF using a gamma distribution. Unlike the normal distribution, the gamma distribution can have skewness of any value, and unlike the log-normal distribution, the gamma distribution guarantees conservation of the mean value between input data distribution and the modeled PDF. Because we calculate the neutrino flux using the mass of U and Th, such conservation of mean is critical. We also pay special attention to detection limits where the values vary largely among researchers. We develop a fitting method that neutralizes the choice of cut-off.

In the last step, we construct 3-D PDFs of U and Th concentrations, $P_{\mathrm{U}}^{(\boldsymbol{x})}(a)$ and $P_{\mathrm{Th}}^{(\boldsymbol{x})}(a)$, by convolving the lithology model $P^{(\boldsymbol{x})}(i|v^{obs})$ with the composition model $f_{\mathrm{U}}^{(i)}(a)$ and $f_{\mathrm{Th}}^{(i)}(a)$ for each location $\boldsymbol{x}$. From the concentration distributions, we calculate the geoneutrino flux as a PDF at Kamioka by summing up the geoneutrino emission rate from each location (which is a straightforward calculation if U and Th concentrations are given) with proper weights depending on the distance to the detector. The PDF sum here needs not only the U and Th concentration probability densities for each location $\boldsymbol{x}$, but also the correlations among all the PDF's at different locations (e.g., correlations among all the combinations of $P_{\mathrm{U/Th}}^{(x_k)}(a)$ and $P_{\mathrm{U/Th}}^{(x_l)}(a)$, for all the $k$ and $l$ location indexes). Neglecting the correlations results in too narrow a summed flux PDF shape, owning to the central limit theorem for uncorrelated variates, but modeling of such correlations is difficult at this time. Here we conservatively assume maximum correlation in order to complete the calculation and we leave the correlation modeling for future work.

## 3 Inference of the lithology-type distribution using a Bayesian approach

### 3.1 Bayesian translator

In many previous studies on deep crustal modeling, various geological insights were employed to obtain lithology models (e.g., Christensen & Mooney, 1995; Rudnick & Gao, 2003). However, the modeling often contains subjective estimation without statistical/theoretical basis, and the uncertainties of the obtained models are not clearly defined. For our goal of constructing probability density functions of the geoneutrino flux at a detector, we need a stronger theoretical framework for quantitative and reproducible error estimation.

In this study, we propose a Bayesian approach to translate seismic velocity in a tomography model into a rock type there. We refer to the results of laboratory experiments by Christensen and Mooney (1995) for P velocities for 29 rock types in the crust. Because they only studied crustal rocks without cracks, we need special treatment for the possibilities that the region is in the mantle or sedimentary layer. Referring to various cross sections obtained by exploration studies in Japan (e.g., Iwasaki et al., 2013), we define the sedimentary layer as the region with P velocity less than 5.0 km/s and the mantle as the region with P velocity greater than 7.5 km/s. For P velocities between 5.0 km/s and 7.5 km/s, we identify the explicit rock type. Hereafter we define the 29 rock types in Christensen and Mooney (1995) as our 1st to 29th rock type, the sediment as the 30th rock type, and the mantle as the 31st rock type.

Here we focus on a location $\boldsymbol{x}$ in the Earth. For notational simplicity, we omit to explicitly write the dependence of $\boldsymbol{x}$. If the P velocity in a tomography model is $v^{obs}$ there, the probability that the lithology is $i$ is evaluated by using the Bayesian theorem:

$$P(i|v^{obs}) = \begin{cases} P(\,5.0 \le v < 7.5 \mid v^{obs})\ \dfrac{P(v^{obs}|i)\,P(i)}{\sum_{j=1}^{29} P(v^{obs}|j)\,P(j)} & (1 \le i \le 29) \\ P(\,v < 5.0 \mid v^{obs}) & (i = 30) \\ P(\,7.5 \le v \mid v^{obs}) & (i = 31) \end{cases}, \qquad (1)$$

where $v$ is the actual P velocity (rather than the observed velocity in a tomography model, $v^{obs}$) at the location $\boldsymbol{x}$, $P(v|v^{obs})$ denotes the conditional probability density of the event that the actual P velocity is $v$ when the observed velocity in the tomography model is $v^{obs}$, and $P(v^{obs}|i)$ denotes the conditional probability density of the event that the observed P velocity is $v^{obs}$ when the rock type there is $i$. $P(i)$ is the prior probability.

The key point of this translator is the evaluation of the conditional probabilities, $P(v|v^{obs})$ and $P(v^{obs}|i)$. We model $P(v|v^{obs})$ by

$$P(v|v^{obs}) = \frac{1}{\sqrt{2\pi}\sigma^{obs}} \exp\left(-\frac{(v - v^{obs})^2}{2\,{\sigma^{obs}}^2}\right), \qquad (2)$$

where $\sigma^{obs}$ denotes the error of the tomography model. We model $P(v^{obs}|i)$ by

$$P\left(v^{obs}|i\right) = \frac{1}{\sqrt{2\pi}\sqrt{\,{\sigma^{obs}}^2 + {\alpha^{(i)}}^2{\sigma^{temp}}^2 + {\sigma^{(i)}}^2}} \exp\left(-\frac{\left(v^{obs} - v^{(i)}\right)^2}{2\left(\,{\sigma^{obs}}^2 + {\alpha^{(i)}}^2{\sigma^{temp}}^2 + {\sigma^{(i)}}^2\right)}\right), \qquad (3)$$

where $v^{(i)}$ and $\sigma^{(i)}$ denotes the mean and the standard deviation, respectively, of the measured velocities in the laboratory experiments for the $i$-th rock type $(1 \leq i \leq 29)$, and $\sigma^{temp}$ and $\alpha^{(i)}$ denote the error of the temperature model and temperature coefficients that we use in the interpretation. While $v^{(i)}$, $\sigma^{(i)}$ and $\alpha^{(i)}$ are explicitly presented or easily evaluated from the data in Christensen and Mooney (1995), we need to appropriately evaluate $\sigma^{obs}$ and $\sigma^{temp}$, which are discussed in the next subsection. In this study, we do not consider mixure of lithology in a cell. In Appendix A1 we show that this method can be naturally extended for mixed lithology, and in Section 6 we discuss the effects of the mixture.

## 3.2 Explicit implementations

We first established the prior probability model on the basis of the lithology of the Hidaka metamorphic belt representing an obducted crust of the Kuril arc (Kimura, 1986; Komatsu et al., 1989; Osanai et al., 1991). The belt consists mainly of metamorphic rocks up to granulite-facies and mafic to felsic intrusives with a minor amount of ultramafic rocks. The metamorphic grade increases westward from very low-grade metasedimentary rocks, through schists, para-gneisses, amphibolites, and mafic-granulites intercalated with para-granulites. The intrusive rocks occur at various crustal levels and include gabbros, diorites, tonalites and granites. Further, the unexposed lowermost crust is considered to exist and comprise garnet-bearing mafic granulites and gabbros (Shimura et al., 2004). To obtain a prior probability, we digitized the idealized geological column of the Hidaka crustal section (Fig. 2 of Shimura et al., 2004). In addition, we re-categorized the rock types of the Hidaka metamorphic belt to be consistent with those of Chirstensen and Mooney (1995) so that we can fully utilize their laboratory experiment data that are among the most essential information for lithology identification. The resultant bulk lithology is shown in Table 1, and we used it as the prior probability model, $P(i)$. Note that the prior does not include any depth dependence nor any horizontal geographical location dependence. The depth information is not used despite the vertical pattern apparent in the Hidaka metamorphic belt, because the Japanese crust is severely deformed and therefore the depth distribution at one specific site cannot be used for other locations directly. This is also important for our approach of not providing crustal boundary model in the prior. For simplicity of notation, we hereafter use three-letters abbreviations in Table 1 to specify the rock type. We also use SED and MTL to denote the 30-th (sediment) and 31-st (mantle) rock type.

The basic geophysical data of our Bayesian translator is a tomography model. In this study, we use the P wave model by Matsubara and Obara (2011). This is one of the highest resolution tomography models obtained by an inversion of the delay times from the modern and dense seismograph network (Hi-net) in Japan. They also obtained the S wave model, however we do not use it this time because the quality of P and S models can be completely different and apparent P and S velocity ratios can have unacceptably large uncertainties. The model is represented using discrete grids with depth intervals of 2.5-10 km and horizontal intervals of 0.1 degrees. We translate the P velocity on each grid point, $v^{obs}$, to the rock type. We need to evaluate the errors of the tomography model, $\sigma^{obs}$, which is a difficult task because this quantity was not explicitly evaluated. Matsubara and Obara (2011) assumed random errors with a standard deviation of 0.05 s in their travel time data, and conducted a recovery test using a checkerboard with a scale of 20 km and heterogeneities of peak amplitudes of $\pm 5\%$ (Fig. 2 of Matsubara and Obara, 2011). One difficulty is that the uncertainties depend on the scale length and the amplitudes of the heterogeneities. However, we can confirm that the dominant scale length and r.m.s. amplitudes of heterogeneities are comparable between the output model of the resolution test and resultant tomography model (see Fig. 2 of Matsubara and Obara, 2011). We therefore assume that the residuals of the input and output velocities of their recovery test are equal to $\sigma^{obs}$. Note that we evaluate $\sigma^{obs}$ as a function of location $\boldsymbol{x}$ because the errors are expected to be highly heterogeneous in space. The error estimation presented above considered all of the important factors to control errors in the tomography model, nonlinearity, observational errors, and amplitude and scale-length of heterogeneities. This is the best option that we can take at present. Our primary purpose here is to develop a methodological framework to stochastically infer lithology from input tomography, hence we do not develop a new tomography model so as not to blur the main point.

In translating the observed P velocity data to rock type, we refer to the laboratory measurements of P-velocities of each rock type by Christensen and Mooney (1995). They presented the average and the standard deviation of measured P velocities for each rock type. They measured at room temperature, but they also presented velocities for three different temperature-depth profiles with extrapolation based on appropriate temperature and pressure coefficients. The profiles used here are those by Blackwell (1971) who presented temperature profiles for low, average, and high heat flow regions in the US (see Fig. 12 of Christensen & Mooney, 1995). For our translation, we assume that temperature in the Japan crust is laterally homogeneous and its depth-profile is the same as that for the average heat flow region in the US. We also assume that the error of this temperature model, $\sigma^{temp}$, is equal to the standard deviation of the three temperature profiles in Blackwell (1971). Despite these assumptions, our model appears to be more or less compatible with a temperature structure model of Japan by Furukawa (1995), in which the temperature structure at 30 km depth (Fig. 5 of Furukawa, 1995) distributes roughly between 150 °C and 850 °C with the average at around 500 °C, while Blackwell (1971) assumes the temperatures at 30 km depth to be 291 °C, 467 °C, and 786 °C in their three temperature profiles. Considering that $\sigma^{temp}$ is small compared with $\sigma^{obs}$ and $\sigma^{(i)}$ (shown in Fig. 2b), it is unlikely that the above assumptions change the main conclusions. For the same reason as the tomography model, we did not develop a new temperature model in this paper.

In this study, we define our region of interest as the lands of the Japan arc and their surrounding continental margins. In this way the area without sufficient seismic ray coverage is excluded, where the error in the tomography model in such area is primarily controlled by the a priori information whose plausibility is not always clear enough in the study of Matsubara and Obara (2011) and most of other tomography studies. Among the grid nodes defined in the model of Matsubara and Obara (2011) with 2.5-10 km vertical intervals and 0.1°×0.1° horizontal intervals (roughly 6.4 km EW × 11 km NS), we selected the nodes in the region of interest as follows. For the depth extent, we select the nodes between the 0 km and 50 km depths. For the horizontal extent, we first define the longitude and latitude ranges as: longitude between 129°E and 146°E, and latitude between 30°N and 45.5°N, excluding latitude > longitude - 94.6° to remove Korea and Russia. Then we select the nodes with elevations above zero (land nodes) and sea-floor depth less than 200 m (continental margin nodes). In order to smooth the boundary, we added nodes adjacent to the land and continental margin nodes. Finally, we excluded all nodes that are more than three nodes apart from their closest land node, to ensure proper coverage by the Japanese seismic observation network. A map of selected nodes are given in Fig. A3 in Appendix. For each selected node, we infer lithology at the location of the node. To fill the volume, cells are defined in such a way that each cell contains one node at its center, except for the surface cells which are trimmed according to topography. A 3-D probabilistic chemical concentration map is then constructed based on the cells.

## 3.3 Results

Fig. 2a shows the expected P velocities (i.e., the velocities presented by Christensen and Mooney (1995) for the temperature profile of the average heat flow region) for the rock types that have non-zero probability in the prior model. For later discussions, we classify the rock types into four categories: uppermost (shown in light blue), upper (shown in red), middle (shown in purple), and lower crustal rocks (shown in blue).

Fig. 2b shows $\sigma^{(i)}$ presented in Christensen and Mooney (1995), $\sigma^{obs}$ evaluated from the recovery test in Matsubara and Obara (2011), and $\sigma^{temp}$ evaluated from the temperature profiles of Blackwell (1971). $\sigma^{obs}$ is laterally heterogeneous and the heterogeneity increases with depth. $\sigma^{obs}$ in Fig. 2b is the averaged value for each depth, and the lateral heterogeneity is shown in Fig A1 for each depth. In the left panel of Fig. 2b, we can identify two rock types that have larger $\sigma^{(i)}$ than others: MGW and QSC. Except for these uppermost and upper crustal rocks, $\sigma^{(i)}$ are, more or less, comparable with $\sigma^{obs}$. In contrast, $\alpha^{(i)}\sigma^{temp}$ is smaller compared with the other two for the uppermost 35 km. Considering that the crustal thickness is less than 35 km for most of the Japan arc (e.g., Iwasaki et al. 2013), we can expect that $\alpha^{(i)}\sigma^{temp}$ does not strongly affect the posterior probability $P^{(\boldsymbol{x})}(i|v^{obs})$.

Comparing Fig. 2a and 2b, we see that $\sigma^{(i)}$ and the other uncertainties are much smaller than the velocity differences among different categories but larger than or comparable with the velocity differences among the rock types classified to the same category. We therefore expect that we can clearly distinguish the categories by the observed seismic velocities, but we cannot clearly distinguish the rock types in the same category unless we refer to information in the prior probability model.

Fig. A2 in Appendix shows the resultant posterior probability model $P^{(\boldsymbol{x})}(i|v^{obs})$ at the 0, 10, 20, 30, 40, and 50 km depths. To show the overall features, we average $P^{(\boldsymbol{x})}(i|v^{obs})$ over the horizontal plane to obtain the relative frequency distribution at each depth (Fig. 3a). Although we did not include any information on the depth in the prior probability, the results clearly show depth preference for each rock type, indicating that we successfully reconstruct the depth distribution using seismological data. Fig. 3b shows the probabilistic map of BGN (upper crustal rock) and GRA (lower crustal rock) at the 20 km depth together with the P-velocity map of the tomography model. We can confirm the tendency that BGN and GRA are likely to distribute in higher and lower velocity regions, respectively. The results show that we also succeed in reconstructing the horizontal geographical distribution using the seismological data.

As another crosscheck, we computed the synthetic Bouguer gravity anomalies from the density distribution obtained from our posterior model, and compared them with the observed Bouguer anomalies (Fig. 4). Here the density distribution is calculated by $\sum_{\mathrm{i}} P^{(\boldsymbol{x})}(i|v^{obs})\,\rho^{(i)}$, where $\rho^{(i)}$ is the rock density for each rock type provided by Christensen and Mooney (1995). The Bouguer anomalies are surface gravity anomalies with corrections from surface topography effects and are approximately proportional to the mass excess or deficiency beneath that point (see Eq. 5-110 of Turcotte and Schubert, 2002). We compute the average density between 0 km and 50 km depths as an index of the mass excess or deficiency. The observed Bouguer anomalies are taken from Shichi and Yamatoto (2001). As the Bouguer anomalies primarily reflect the Moho undulations, the coincidence of overall amplitudes and patterns in Fig. 4 shows that the Moho topography is successfully reconstructed. Since we used a smooth tomography model without explicit crustal thickness data, this result demonstrates that we succeed in reconstructing the vertical distribution as well.

Fig. 5 shows a comparison of the relative frequency distribution in the obtained posterior model ensemble and the prior probability model. Both models show similar proportions, which indicates the plausibility of our choice of the prior model. The coincidence is not trivial because we discarded depth information in developing the prior. The results suggest that the proportions of upper, middle, and lower crustal rocks in the Hidaka metamorphic belt are similar to those in the bulk Japan crust. Fig. A4 in Appendix shows a consequence of using a non-local model for comparison.

## 4 Geochemical modeling using probability distribution functions

### 4.1 Geochemical database for the modeling

Detailed geological and geochemical information nearby the detector is necessary for precise estimation of the regional contribution to the geoneutrino flux. It is known that the composition of the continental crust is heterogeneous and depends on the tectonic setting (e.g., Hacker et al., 2015). Indeed, Togashi et al. (2000) pointed out that the upper crust of the Japan arc has slightly lower U and Th concentrations as compared to typical continental upper crust. As such, in order to evaluate U and Th distributions in the Japan crust, we construct a database of chemical compositions of rocks from the Japan arc rather than using global average values or estimation based on typical continental crust.

We compiled $SiO_2$ wt%, U ppm, and Th ppm of rocks from various locations in the Japan arc (Fig. 6) together with the lithological information. The data are from 96 papers and reports, and cover various rock types such as igneous volcanic rocks, plutonic rocks, sea floor altered oceanic crust, sedimentary rocks, ultramafic rocks, and metamorphic rocks. In order to minimize analytical uncertainties, compositional data are mostly compiled from papers published after 2000 CE with few exceptions. In order to cover numerous rock types of the Japan arc with a sufficient amount of data, some compositional data are taken from studies in the late 90's. To minimize analytical uncertainties, U and Th concentrations obtained by Inductively Coupled Plasma Mass Spectrometer (ICP-MS) and Instrumental Neutron Activation Analysis (INAA) are exclusively used, while the concentrations obtained by X-ray fluorescence (XRF) are excluded from the following calculations. As a result, 2088 analyses of U concentrations and 2182 analyses of Th concentrations are compiled. Data sources, rock description, sample location and analytical method are summarized in Appendix. Some compositional data such as those for volcanic rocks and limestone are not used to estimate the neutrino flux at Kamioka because these rocks do not appear in the prior model constructed from the Hidaka crustal section. In reality, limestone as well as serpentine and quartzite sporadically occurs within accretionary complexes of the Japan arc. Note, however, that its contribution to the neutrino flux should be trivial due to its minor abundance and low concentrations of U and Th (supplementary table S1). By contrast, volcanic rocks are widely distributed near the surface of the arc. As surficial volcanic rocks have P velocities less than 5.0 km/s (Christensen & Mooney, 1995), they should be categorized as the "sedimentary layer" in our lithology inference (Section 3.1). Besides, compositional data for sedimentary rocks would represent the average compositions of eroded crust including volcanic rocks (e.g., Togashi et al., 2000). As a consequence, it is expected that the estimated contribution of the sedimentary layer to the neutrino flux actually involves that of volcanic rocks.

There is a restricted amount of data for the lower crustal rocks and mantle rocks in the compiled database. To cover the shortage, gabbroic xenoliths are newly analyzed for major and trace elements. Xenolith samples are taken from the back-arc regions of the Japan arc; Oki-Dogo in southwest Japan, Ichinomegata in Northeast Japan, and Takashima in Kyusyu island (Fig. 6). Major element concentrations are analyzed by XRF, and trace elements are analyzed using laser-ablation ICP-MS. The analyzed $SiO_2$, U, and Th concentrations, as well as detailed analytical methods, are summarized in Appendix A2.

Compositions of mantle rocks are taken from xenolith data in the global GEOROC database (http:// georoc.mpch-mainz.gwdg.de/georoc/, accessed August 2017). Mantle xenoliths from various settings such as arc, continental, and shield are included in our database, with data from papers published after 2000 CE. Chemical compositions of garnet-bearing peridotite are excluded because we focus on the pressure range that corresponds to the stability of garnet absent lherzolite (<50 km depth). Garnet bearing peridotite may represent a fertile, deep mantle. Altered or metasomatized samples are also excluded from the database to avoid possible modifications by melt, fluid, or surface processes. Although we have carefully selected the mantle rock data, the choice of geochemical data for mantle rocks does not affect the geoneutrino flux calculation for two reasons: 1) the volume of mantle in our region of interest is small, and 2) mantle rocks exhibit very low concentrations in U and Th.

## 4.2 Geochemical rock type classification

Rocks in our database are classified mostly following the scheme of Christensen and Mooney (1995) based on petrological information. However, detailed petrological or structural descriptions such as grain size, foliation, or lineation are unavailable in some of the original papers. For instance, some rocks are described solely based on metamorphic classification without petrological or rock structural descriptions (e.g., "granulite facies" and "granulite"). Therefore, rock type classification based on petrography is sometimes difficult (e.g., metagraywacke, slate or phyllite, and gabbro or garnet-absent granulite). Some different rock types classified in Christensen and Mooney (1995) have clearly distinctive seismic velocities but exhibit no systematic difference in chemical compositions (e.g., granite and granitic gneiss). Accordingly, some rock types in Christensen and Mooney (1995) are combined into larger groups for the purpose of geochemical modeling. Our rock type classification and criteria are summarized in Table 2. Rocks are basically classified based on lithological explanations and mineral assemblages, with the exception of GAB/MGR/GGR, which are distinguished according to $SiO_2$ concentration. Gabbro and granulites with $SiO_2$ < 52 wt% are classified as GAB/MGR/GGR, following the International Union of Geological Sciences (IUGS) classification of igneous rocks (Le Bas & Streckeisen, 1991). We consider this criterion to be reasonable for U and Th modeling because U and Th concentrations exhibit noticeable stepwise increases around $SiO_2$ = 52 wt% as shown in Fig. 7. The stepwise increase indicates that concentration processes of U and Th for rocks with $SiO_2$ above and below 52 wt% are different.

Based on the above classifications, we constructed a compositional database for each rock type used for the geochemical modeling, which is presented in the supplementary material. Scatter plots of $SiO_2$, U, and Th concentrations are shown in Fig. 7. The number of samples, means, and medians of U and Th concentrations of each rock type are given in Table 3. Locations of samples in the database are summarized in Fig. 6 and in the supplementary material.

### 4.3 Composition PDF construction

#### 4.3.1 PDF models and the Gamma distribution

U and Th compositions of each rock type are modeled as probability distribution functions (PDF). Some necessary conditions for the composition PDF are: (1) being able to adapt to both symmetric (Gaussian-like) and asymmetric (skewed) distributions, (2) zero-probability for negative concentration values, and (3) conservation of mean value upon model fitting. Both the normal (Gaussian) and log-normal distributions fail to satisfy these conditions. Although quantitative definition is difficult, a good fit to actual rock composition distributions is important, and some connection to underlying geological processes is desired.

The normal distribution is the maximum entropy probability distribution when arithmetic mean and deviation are constrained. Likewise, the log-normal distribution is the maximum entropy distribution when geometric mean and deviation on log scale are constrained. When arithmetic mean and geometric mean are constrained, the maximum entropy distribution becomes the gamma distribution. The gamma distribution is a two-parameter continuous probability distribution, defined as

$$f(x;k,\theta) = \mathrm{Gamma}(x;k,\theta) = \frac{x^{k-1}e^{-\frac{x}{\theta}}}{\Gamma(k)\theta^k}, \qquad (4)$$

with shape parameter $k$ (>0) and scale parameter $\theta$ (>0). The skewness is given by $2/\sqrt{k}$, allowing the distribution to be highly skewed (small $k$) or Gaussian-like (large $k$). The probability of negative values is zero, and as discussed below, the model does not bias the mean value. Although the gamma distribution is not the unique solution for composition modeling, it is at least a better choice than the normal or log-normal distributions for our purpose of unbiased flux calculation. In the following, we examine other characteristics of the gamma model.

#### 4.3.2 Parameter fitting

We use the maximum likelihood method (or, equivalently, MAP estimation with a flat prior) to determine the parameters of the model PDF. Let $f(x;\boldsymbol{p})$ be a compositional model PDF (not necessarily a gamma distribution), with $x$ being the concentration random variable and $\boldsymbol{p}$ being the model parameters. In principle, we can estimate the parameters $\boldsymbol{p}$ for a given observation of rock samples $\boldsymbol{a} = \{\, a_1,\ a_2,\ \dots, a_N \,\}$ by minimizing the negative log-likelihood function (with a factor of $-2$ for conventional reasons),

$$-2\ \log\mathcal{L}(\boldsymbol{p};\boldsymbol{a}) = -2\sum_{j=1}^{N}\log f\big(a_j;\boldsymbol{p}\big). \qquad (5)$$

In case of the gamma distribution model, the best-fit PDF, $f(x; \boldsymbol{p}) = \mathrm{Gamma}(x; \hat{k}, \hat{\theta})$, where $\hat{k}$ and $\hat{\theta}$ are the best-fit parameters determined by minimizing the negative log-likelihood, has the expected value exactly equal to the sample mean, regardless to the actual distribution of the samples (details in Appendix A3), i.e.,

$$E\{x\} = \hat{k}\hat{\theta} \equiv \frac{1}{N}\sum_{j=1}^{N} a_j . \qquad (6)$$

Therefore maximum likelihood fitting of a gamma distribution to any collection of samples does not bias the mean value. If the fit gamma model is used for further mass-balance type calculation, such as geoneutrino flux integration and thermal balance, its resultant mean and total will not be biased, even if the gamma model is not correct (i.e., there is no modeling bias).

In general, rock samples can have elemental concentrations below a detection limit, which invalidates the simplistic likelihood function given by Eq. (5). Since the real concentration below the detection limit is unknown, those samples must be removed from the shape-fitting likelihood, but they can still be included in the parameter estimation. For samples above the detection limit, a similar log likelihood function to Eq. (5) can be constructed to fit the distribution shape, with modification to account for the re-normalization due to the partial sample removal. With $B (> 0)$ as a lower-side cutoff (not necessarily the detection limit), it is given by

$$-2\ \log \mathcal{L}_{\mathrm{shape}}(\boldsymbol{p}; \boldsymbol{a}) = -2 \sum_{j=1}^{N} \Theta(a_j - B)\ \log \left\{ \frac{f(a_j; \boldsymbol{p})}{\int_B^{\infty} f(\alpha; \boldsymbol{p})\, d\alpha} \right\}, \qquad (7)$$

where $\Theta(a)$ is the Heaviside step function. For samples below the cutoff, a binominal distribution can be used for the probability of the concentration being below the cutoff. With $N_0$ as the number of samples below the cutoff, it is given by

$$-2\ \log \mathcal{L}_{\mathrm{count}}(\boldsymbol{p}; \boldsymbol{a}) = -2 \log \left\{ \binom{N}{N_0} \left( \int_0^B f(\alpha; \boldsymbol{p}) d\alpha \right)^{N_0} \left( \int_B^{\infty} f(\alpha; \boldsymbol{p}) d\alpha \right)^{N-N_0} \right\}. \qquad (8)$$

Our final log-likelihood is a combination of these two,

$$-2\ \log \mathcal{L}(\boldsymbol{p}; \boldsymbol{a}) = -2\ \log \mathcal{L}_{\mathrm{shape}}(\boldsymbol{p}; \boldsymbol{a})\ - 2\ \log \mathcal{L}_{\mathrm{count}}(\boldsymbol{p}; \boldsymbol{a})\ . \qquad (9)$$

If the model PDF extends to negative values, as in the normal distribution model, the binominal distribution is replaced with a polynomial distribution with three segments: negative values, positive below cutoff, and positive above cutoff. However, setting the number of samples with negative concentration zero reduces it to the formula in Eq. (8).

Although we do not adopt it this time, this method can also be applied to a cutoff for an upper boundary, which will improve modeling robustness against whether or not rare high-concentration samples are found in the first place. This will also help improve robustness against outliers by exotic samples that might have been included due to improper rock type classification.

### 4.3.3 Detection limits and sample cutoff

In the method described above, any value may be chosen for the cutoff as long as it is set above or equal to the detection limit. If the detection limit was fixed, or even just known, we would be able to use its value for optimal sensitivity. But it is often the case that the detection limit is not stated in papers, and the limit varies depending on analytical methods. Here we empirically choose a cutoff value based on the relationship between the U/Th ratio and U concentration, given that crustal rocks normally exhibit a restricted range of U/Th ratio. Fig. 7d shows that U/Th ratio scatters becomes large at lower U concentrations, especially in the range U < 0.2 ppm. While such large scatter could in principle have a geological origin, it is also the case that for a particular rock, Th concentrations are expected to be analyzed more accurately than U due to the generally higher Th concentrations. Hence, the increased scatter in U/Th ratio for low-U samples can be interpreted as possibly resulting from inaccurate determination of U concentration at <0.2 ppm. Considering the possible inaccuracy, we conservatively take 0.2 ppm for our uniform cutoff value. It should be noted that, although we take a uniform value (0.2 ppm) for the cutoff value in all data, the cutoff value can be arbitrary chosen, and can be variable in each rock type, element, or dataset.

### 4.4 Result of the geochemical modeling

The relationships between the $SiO_2$, U and Th concentrations of the rocks used for our geochemical modeling are shown in Fig. 7. Average compositions of upper, middle and lower crust (Rudnick & Gao, 2003), depleted MORB mantle (Workman & Hart, 2005), pyrolite (McDonough & Sun, 1995) and an average composition of the upper crust of the Japan arc (Togashi et al., 2000) are also shown for comparison. GRA/GGN exhibits the highest U and Th concentrations among the rock types used here. BGN exhibits relatively depleted U and Th concentrations compared to those of GRA/GGN at a fixed $SiO_2$ concentration. Sedimentary and metasedimentary rocks (MGW/SLT/PHY, QSC, PGR) exhibit intermediate U and Th concentrations regardless of their $SiO_2$ concentrations, indicating that U and Th concentrations are homogenized during sedimentary processes and surface modification, and that these surface processes are not effective at concentrating U and Th. In our dataset, no systematic compositional difference between xenolith and outcrop granulite samples is observed, in contrast to Rudnick and Presper (1990) who found a difference between terrane and xenolith granulites (Fig. A5 in Appendix).

The mean values of Th and U concentrations of mantle rock in this study are 0.23 ppm and 0.10 ppm, respectively (Table 4). For comparison, Huang et al. (2013) report 0.150 ppm for Th and 0.033 ppm for U as the “mean” composition of lithospheric mantle (Table 5 in Huang et al., 2013). If the raw samples are compared, both datasets basically agree with each other in all the basic statistics of U and Th concentrations, such as compositional ranges (upper and lower limits of concentrations and U/Th ratio), geometric mean, arithmetic mean and median. The discrepancy in reported mean appears to result from the fact that Huang et al., (2013) calculates the mean on a log scale (i.e., the geometric mean) and also applies a 1.15 sigma (in log) filter around the median (= mean in log-normal). As discussed in Section 6, geometric mean in the log-normal model is basically the median value, and for positively skewed distributions, median becomes smaller than the arithmetic mean.

Fig. 8 shows the composition PDF models determined by the maximum likelihood fitting to our dataset, together with histograms of U and Th concentrations in the rock samples. The parameters of the fit PDFs, together with the mean and median values of the fit PDFs, are shown in Table 4. As mentioned above, the gamma distribution model reproduces the mean value of the sample distribution. Even with the 0.2 ppm cutoff, the mean values for >0.2 ppm samples are accurately reproduced by the >0.2 ppm part of the gamma distribution, while biases are observed for both the log-normal and normal distributions.

Table 5 shows the log-likelihood values at the optimal fitting, for the gamma distribution, log-normal distribution and normal distribution models. Note that all these distributions have two parameters and the Akaike Information Criterion (AIC) is equally given by $-2\log\mathcal{L}-2$, therefore the values in Table 5 provide immediate comparison among models. Although log-normal distribution and normal distribution exhibit better fits than the gamma distribution for some rock types, as a whole, the best log-likelihood values in Table 5 indicate that the gamma distribution can be used to model both skewed and symmetric distributions.

If the composition PDFs are sorted in the order of increasing $k$, we obtain MTL < GAB/MGR/GGR < AMP < PGR < BGN < GRA/GGN < MGW/SLT/PHY < QSC for Th, and MTL < GAB/MGR/GGR < AMP < BGN < GRA/GGN < PGR < MGW/SLT/PHY < QSC for U. Mafic rocks exhibit lower *k* values, and felsic rocks and sedimentary rocks exhibit higher *k* values. This means mafic rocks exhibit high skewness and felsic and sedimentary rocks exhibit symmetrical distributions.

## 5 Geoneutrino flux

### 5.1 Flux calculation from given source distributions

If U and Th distributions are given, the geoneutrino flux $\Phi$ at detector location $\boldsymbol{x}_{\mathrm{d}}$ can be calculated by simply summing up the neutrino emission rate from each location $\boldsymbol{x}$, weighting for solid angle:

$$\Phi_{\mathrm{U/Th}} = A_{\mathrm{U/Th}} \int \frac{a_{\mathrm{U/Th}}(\boldsymbol{x})\,\rho(\boldsymbol{x})}{4\pi|\boldsymbol{x}-\boldsymbol{x}_{\mathrm{d}}|^2}\, d^3\boldsymbol{x}, \qquad (10)$$

where $A$ is the neutrino emission rate from the $^{238}$U and $^{232}$Th decay chains, per natural U and Th elemental mass, $7.410\times10^{7}$ (kg$^{-1}$s$^{-1}$) and $1.623\times10^{7}$ (kg$^{-1}$s$^{-1}$), respectively (Enomoto 2005), $a_{\mathrm{U/Th}}(\boldsymbol{x})$ represents the U and Th concentrations, and $\rho(\boldsymbol{x})$ is the rock density at location $\boldsymbol{x}$. Neutrinos change flavor during travel, modifying the fraction of the flux detectable by the inverse beta-decay process, which is currently the only practical way to detect geoneutrinos. Although the detectable fraction depends on the energy of the neutrinos and the distance traveled, for the case of geoneutrinos the integration averages out those dependences, and the effective detectable fraction can be approximated at better than 1% accuracy by a single factor (Enomoto 2005) as:

$$\frac{\Phi^{\mathrm{detectable}}}{\Phi} \sim \cos^4\theta_{13}\left(1-\frac{1}{2}\sin^2 2\theta_{12}\right) + \sin^4\theta_{13} \sim 1-\frac{1}{2}\sin^2 2\theta_{12}, \qquad (11)$$

where $\theta_{12}$ and $\theta_{13}$ are neutrino oscillation parameters that have been determined by oscillation experiments, including KamLAND itself. One should refer to the latest measurements of the neutrino oscillation parameters from the literature to convert from the total flux discussed here to the observable flux. For example, the Particle Data Group 2016 (Patrignani et al., 2016) gives $\sin^2\theta_{12} = 0.304^{+0.014}_{-0.013}$, which leads to $\Phi^{\text{detectable}}/\Phi = 0.577 \pm 0.011$. In some works, the total flux $\Phi$ is referred to as the "no-oscillation flux".

## 5.2 Flux probability model

The lithology map constructed in Section 3 provides for each location $\boldsymbol{x}$ the probability $P^{(\boldsymbol{x})}(i)$ of the rock type being $i$, for all the rock types defined in the prior distribution. Christensen and Mooney (1995) provides the rock density $\rho(i,d)$ of rock type $i$ at depth $d$ for various depths. The PDF of U and Th concentrations $a_{\text{U/Th}}$ at location $\boldsymbol{x}$, $g^{(\boldsymbol{x})}_{\text{U/Th}}(a)$, are calculated for each location $\boldsymbol{x}$ by convolving the lithology PDFs $P^{(\boldsymbol{x})}(i)$ and the U and Th composition PDF $f_{\text{U/Th}}(a|i)$ for given rock type $i$ (denoted as $f^{(i)}_{\text{U/Th}}(a)$ in Section 4),

$$g^{(\boldsymbol{x})}_{\text{U/Th}}(a) = \sum_i f_{\text{U/Th}}(a|i) \cdot P^{(\boldsymbol{x})}(i). \qquad (12)$$

From the U and Th distribution PDF at location $\boldsymbol{x}$, $g^{(\boldsymbol{x})}_{\text{U/Th}}(a)$, the flux PDF is calculated by integrating the neutrino emission from each location. As the source distribution is expressed by a PDF, the result of integration also becomes a PDF, $h_{\text{U/Th}}(\Phi)$. The expected value of the flux can be easily calculated by replacing all the PDFs with their expected values,

$$E\{\Phi_{\text{U/Th}}\} = A_{\text{U/Th}} \int \frac{E\{a^{(\boldsymbol{x})}_{\text{U/Th}}\}\,\rho(\boldsymbol{x})}{4\pi|\boldsymbol{x}-\boldsymbol{x}_{\text{d}}|^2}\, d^3\boldsymbol{x}. \qquad (13)$$

Redefining $E\{\Phi_{\text{U/Th}}\}$ as $\Phi_{\text{U/Th}}$ and $E\{a^{(\boldsymbol{x})}_{\text{U/Th}}\}$ as $a_{\text{U/Th}}(\boldsymbol{x})$ reduces the integration to Eq. (10), which corresponds to a flux calculation without error estimation.

When dealing with multiple PDFs, Monte-Carlo (MC) is a common (and probably the only, if PDFs are of arbitrary distributions) method. Here samples from $g^{(\boldsymbol{x})}_{\text{U/Th}}(a)$, denoted as $\alpha^{(k)}_{\text{U/Th}}(\boldsymbol{x})$ with sample index $k$, are randomly generated according to the PDF shape, and for each sample, the integration is calculated by

$$\varphi^{(k)}_{\text{U/Th}} = A_{\text{U/Th}} \int \frac{\alpha^{(k)}_{\text{U/Th}}(\boldsymbol{x})\,\rho(\boldsymbol{x})}{4\pi|\boldsymbol{x}-\boldsymbol{x}_{\text{d}}|^2}\, d^3\boldsymbol{x}. \qquad (14)$$

By repeating the sampling and integration a number of times, typically thousands to millions, the frequency distribution of $\varphi^{(k)}_{\text{U/Th}}$ reveals the shape of the $h_{\text{U/Th}}(\Phi)$ probability distribution.

In this MC method, the central problem is the sampling of $\alpha^{(k)}_{\text{U/Th}}(\boldsymbol{x})$ from the PDF $g^{(\boldsymbol{x})}_{\text{U/Th}}(a)$. If correlations exist among different locations, which is likely the case, sampling must be done simultaneously (or at least coherently, as typically done in the Markov Chain Monte Carlo technique) from all locations in order to preserve the correlations. Since $g^{(\boldsymbol{x})}_{\text{U/Th}}(a)$ consists of the lithology model $P^{(\boldsymbol{x})}(i|v)$ and the composition model $f^{(i)}_{\mathrm{U/Th}}(a)$, various correlations can exist, such as lithological similarities between neighboring cells, compositional similarities within each geological unit, and the well-known correlation between U and Th concentrations. Neglecting correlations in PDF integration results in too narrow distributions as the central limit theorem shows for the sum of independent random variates. Too narrow a distribution (too small estimated uncertainties) leads to unsound conclusions deduced from the inputs.

Despite the importance of correlations, their proper modeling in the lithological and compositional models is not straightforward. We leave this topic for future work. For now, being conservative, we perform the PDF integration assuming maximum correlations among all values. This will result in an overestimate of uncertainty in the final flux estimation, but at least the overall method still achieves our goal of constructing a fully probabilistic objective modeling.

Under maximum correlation, if one random variable is at its minimum (center, or maximum), then all the other variables must be at their minimum (center, or maximum). We construct the maximum correlation by having only one global random variable $q_k$ for each $\varphi_k$, and use it as a quantile of each PDF, $P_{\mathrm{U/Th}}(a, \boldsymbol{x})$. Here $q_k$ distributes between 0 and 1 uniformly. In this way, any two PDFs for different locations $\boldsymbol{x_1}$ and $\boldsymbol{x_2}$ or for different radioactive elements U and Th are fully correlated due to the single shared random variable, yet for each individual PDF the samples generated out of it still follow the PDF shape. (In this case there is only one random variable uniformly distributing between 0 and 1, and there is no need to actually generate random numbers: a uniform scan from 0 to 1 will produce the same result. For this reason, the flux integration here is no longer a true MC.)

### 5.3 Flux calculation result

Fig. 9 shows the result of flux integration for the geoneutrino fluxes at the KamLAND location (36.42°N, 137.31°E) from $^{238}$U and $^{232}$Th decays within the region of interest (Fig. A3 in Appendix). The histograms are frequency distributions of $\varphi^{(k)}_{\mathrm{U}}$ and $\varphi^{(k)}_{\mathrm{Th}}$ calculated using Eq. (14). The mean values of the distributions are $1.89 \times 10^6$ (cm$^{-2}$ s$^{-1}$) and $1.86 \times 10^6$ (cm$^{-2}$ s$^{-1}$) for U and Th, respectively. As expected, these basically agree with the mean values from a direct calculation using Eq. (13), $1.87 \times 10^6$ (cm$^{-2}$ s$^{-1}$) and $1.86 \times 10^6$ (cm$^{-2}$ s$^{-1}$) for U and Th, respectively. The RMS spreads of the histograms, as a measure of the flux estimation error, are 61% to 69% of the central values. These large errors are a direct result of the maximum correlation assumption, as discussed previously. If all the correlations were neglected in this calculation, the RMS of the histograms would shrink to an unphysically small value of ~3%, demonstrating the importance of proper correlation modeling. We will pursue improved correlation modeling in future work.

For comparison with past models, the geoneutrino flux at Kamioka from the same region-of-interest is also plotted in Fig. 9. The plot is calculated using the crustal model in Enomoto et al. (2007) with minor updates, which uses CRUST 2.0 (Laske et al., 2001) for crustal boundaries and Rudnick and Gao (2003) for crustal composition. (The original Enomoto et al. (2007) used an old crustal composition model by Rudnick and Fountain (1995); the update alters the flux by -5.7% and +0.3% for U and Th, respectively.) The calculated fluxes, $1.78 \times 10^6$ (cm$^{-2}$ s$^{-1}$) for U and $1.56 \times 10^6$ (cm$^{-2}$ s$^{-1}$) for Th, are also shown in Fig. 9. To combine the local contribution model in this work with a global model, one has to subtract these numbers from the global flux calculation.

For comparison of uncertainties with past models, Table 6 summarizes the methods and estimated uncertainties (the values in the last column are discussed in Section 6). Since previous models do not have probabilistic uncertainty estimations, quantitative comparison cannot be made.

## 6 Discussion

In this study, we perform a lithology inference, taking the regional characteristics of lithology into account. Indeed, the resultant lithology is somewhat different from that in the typical continental crust (e.g., Christensen & Mooney, 1995), in particular a significant amount of sedimentary (MGW, SLT) and metasedimentary rocks (PGR) is observed (Fig. 3a). Because Japan was located on the peril of the land mass of Eurasia, a large amount of sediments should have been accumulated. Indeed, from the extensive analysis of surface geology, Togashi et al. (2000) confirmed that proportions of sedimentary rocks at the surface are significantly larger in Japan than in older continents. We also infer U and Th geographical distributions taking the regional characteristics of chemical composition into account. Slightly lower U and Th concentrations in the upper crust (Togashi et al., 2000) are also manifested in our model (Fig. 10).

Our lithology inference does not depend on crustal segmentation into upper, middle and lower regions. The resultant lithological and chemical models (e.g., Fig. 3a and 10) show gradual change with depth. Besides the fact that the definitions of each segment are somewhat arbitrary, developing segmented models often requires slightly *ad hoc* interpolation or extrapolation (Laske et al., 2001; Coltorti et al., 2011). Our method does not require such manipulations, and allows for the development of fully data-based models. Our composition modeling depends neither on segmentation (upper/middle/lower crust) nor on binary felsic/mafic classifications. Our modeling makes use of the fact that BGN has lower U and Th concentrations compared with GRA/GGA with similar $SiO_2$ concentrations.

One of the limitations of our lithology inference is that we did not take the effects of melt into account. Because the presence of melt in the crust strongly affects the S velocity, the use of such information is an important future research topic. We need an appropriate lithology translator that takes melt effects into account. Ueki and Iwamori (2016) already proposed such a method, and we plan to apply their method in the near future. The use of the information of electric conductivity is also an important future research topic. We also plan to apply the method of Pommier et al. (2008) to interpret observed conductivity anomalies of hydrous and dry melts.

In our lithology PDF model we assume that each cell consists of single lithology. In Appendix A1 we show that our framework can be naturally extended to mixed lithology, with an example of two-lithology mixture. Fig. A6 in Appendix compares the inferred models from the single lithology model and from the two component mixure. The results show that the depth distribution of each rock-type is slightly blurred with the mixture, but the overall patterns remain unchanged. The posterior bulk lithology composition shows slight increase of the two edge components, MGW (slowest) and GGR (fastest), as a consequence of general broadening, but it is basically the only difference. The geoneutrino flux PDFs calculated with the mixed lithology model differ from the single lithology model only by 3% (U) and 4% (Th) in the mean values, with increase in uncertainties from 60% to 70% (U) and 68% to 76% (Th) reflecting the broader lithology distributions. The comparison shows that this additional freedom only makes the result broader by ~15%, without significantly altering the structure.

In our geochemical modeling, rock compositions are described by PDFs, and by adopting gamma distributions for the PDFs, the mean values are conserved and thus the final flux is not biased. This is mathematically shown in Appendix A3 and also verified quantitatively in Tables 3 and 4. Log-normal models do not guarantee such conservation, but one might argue that log-normal models have been used for various composition modeling, and even without guarantee, the behavior in general appears to have been satisfactory. In the case we encountered, a large bias arising from the log-normal model was suspected due to a problem in the data, such as too small a number of samples. To provide an example, if a log-normal model is used to fit to our U concentration data of basalt (n=496, data available in the supplementary material), the expected value of the fit PDF, 0.388 ppm, is ~10% higher than the sample mean value, 0.351 ppm. For other rock types with smaller n, we observe differences of more than one order of magnitude. This leads us to seek for a mechanism to guarantee mean conservation.

If “mean” is calculated on a logarithmic scale (i.e., calculating geometric mean, as opposed to arithmetic mean), one needs to pay special attention, as observed in the discrepancy in mantle rock compositions between this study and Huang et al. (2013) (Section 4.4). Since mean is equal to median for the normal distribution, the geometric mean becomes equal to median if the distribution is log-normal, and therefore the “mean” in Huang et al., (2013) is actually the median of their modeled distribution. In fact, in their Table 5, the filtered mean and unfiltered median are exactly the same values for U and Th (and are consistent with our median values). For depleted rocks (i.e., MTL, GAB/MGR/GGR, AMP in Fig. 8) with highly skewed distributions, the median may become one order of magnitude smaller than the mean. While the median can be a good statistic for describing rock characteristics and for classification, if it is used for mass-balance calculations the result could be biased. For example, U and Th concentrations of the mantle are particularly important for conducting forward calculations of mantle thermal evolution (e.g., Honda, 1995) and geochemical mass balance evolution (e.g., Paul et al., 2002). The "representative compositions" should be carefully chosen in conducting such geochemical forward modeling.

By using the method described in Section 4.3.2, we can properly include the samples below the detection limit and also we can safely apply cuts to exclude outliers without causing bias, eliminating the main reason for using the median. However, if samples are already biased at the time of collection, the bias will be propagated to the final result. While biases already in the input datasets are beyond our scope, the impact of sampling biases to the geoneutrino results is estimated in Appendix A4. Note that median and log-normal models are also susceptible to the sampling bias in the same way.

Fig. 10 shows the vertical distributions of U and Th calculated from our lithological and compositional models for the Japan crust area (our region-of-interest, Fig. S2). For the cells in each depth layer, concentrations are averaged under the maximum correlation assumption. For comparison, the Fig. also shows the crustal structure model in CRUST 2.0 (Laske et al., 2001) and the global average crustal compositional model by Rudnick and Gao (2003). The geoneutrino fluxes calculated by those existing crustal models, shown in Fig. 9 as dotted lines, are in fact very close to the mean values of our flux PDFs. It is encouraging to see the agreement despite the large uncertainties in our model, but we emphasize that comparison with values that do not have proper uncertainties cannot be performed quantitatively in the statistical sense. Our primary goal is to make flux estimation with truly probabilistic uncertainties.

The large error in our result is primarily due to the maximum correlation assumption. In particular, this assumption directly projects the broad distributions of rock composition PDF (Fig. 8) to the final flux PDF (Fig. 9). However, situations will be much improved if spatial correlations are weak for larger length scales. The composition PDF basically represents the spread of U and Th concentrations in rock samples, which are typically a few cubic centimeters in size. It is often observed that rock samples of one type in a single geological unit show quite a broad spread in U and Th concentrations. Sometimes the spread is large enough to fully cover the entire composition PDF, even if all the samples are taken from one single geological unit (typically of a size of ~km). If this is the case, the composition PDF constructed from "~km size samples" would have much narrower distributions, due to averaging out of small-piece fluctuations.

To speculate about the effect of the averaging of small-scale fluctuations, if our composition PDFs were replaced with Gaussian PDFs with mean set to the sample value mean and sigma set to error of the mean estimation (as opposed to the spread of sample values), then the final flux uncertainties would be reduced to 25% ~ 35%. Although this cannot be justified without quantitative evaluation, it should be noted that this treatment basically corresponds to the methods used in past geoneutrino calculations. For example, the errors quoted in Rudnick and Gao (2003) are basically the errors of central value estimations, not the spreads of the values, and these errors are directly propagated to the final flux errors in, e.g., Huang et al. (2013) and Šrámek et al. (2016). In this treatment, possible regional variations might have been underestimated. We are constructing a new rock composition dataset with accurate location information (Haraguchi et al., accepted), and with that, we expect to be able to (at least partially) make quantitative estimation on the spatial correlations.

The local crust region contributes ~50% of the total geoneutrino flux observed at Kamioka (Enomoto et al., 2007). Therefore our crustal model flux errors, 61% for U and 69% for Th, are translated to ~33% error to the total flux. If the averaging discussed above is assumed, this becomes ~15% error in the total flux. On the other hand, the total radiogenic heat from U and Th decays in the earth is predicted in the range of 7.5 TW ("Low-Q" model) to 31 TW ("High-Q" model) among commonly used earth models in recent geoneutrino studies (e.g., Šrámek et al., 2016 and Gando et al., 2013). Considering models of U and Th distributions in the mantle, ~60% of flux variation at Kamioka is expected depending on earth models. Although further reduction of our crustal model uncertainty is desired, our current crustal composition PDF can still be used to put useful constraints on the earth models, and describing crustal model errors as PDF will provide with a statistically proper way to combine earth models and experimental observations.

Regarding the sample-size dependence, the gamma distribution composition model has interesting characteristics. Namely, if concentrations of small-piece samples follow a gamma distribution, then concentrations of imaginary large samples (which is a combination of the small piece samples) also follow a gamma distribution, owing to the reproductive property of the gamma distribution. For example, if U concentrations of small-piece samples follow Gamma($k$, $\theta$), where $k$ is the shape parameter and $\theta$ is the scale parameter, the U concentrations of large-piece samples created by the combination of $N$ independent random small-piece samples follow Gamma($Nk$, $\theta/N$). Note that the skewness of the gamma distribution is given by $2/\sqrt{k}$, making the large-piece distribution less skewed than the small-piece distribution. The factor $1/N$ of the scale parameter keeps the average concentration constant. This characteristic makes the gamma model universal against sample-size dependence, under the approximation of no correlation among samples (large variation in samples, compared to inter-regional variations in mean compositions). Fig. A7 in Appendix shows PDF shapes of gamma distributions with several $k$ values.

This reproductive property also appears among sums of gamma distribution variates with distinct $k$ parameters (the resultant gamma distribution's $k$ is the sum of the input $k$'s). Even if both $k$ and $\theta$ parameters are distinctive, the distribution of the sum can also be approximated by a gamma distribution (an example is shown in Fig. A8 in Appendix; precisely, it is a series of gammas (Mathai, 1982; Moschopoulos, 1985)). As an example from a geological context, consider the observation that mafic rocks generally have highly skewed distributions (small $k$) and sedimentary rocks generally have Gaussian-like distributions (large $k$). If a mixing process is dominantly responsible for the compositional development of rocks from mafic igneous to felsic igneous to sedimentary, then it explains why the gamma model can fit well to all the different rock types (In this example, the scaling the parameter $\theta$ corresponds to the concentrating of trace elements). In fact, if rock types are sorted by the order $k$ in our model, they basically trace the petrogenetic cycle pattern (Fig. 8 already sorts the rocks in this order). We also remark that if $k$ is small, the distribution is exponential ($k = 1$) or close to inverse-proportional ($k \sim 0$), which is a solution of the diffusion equation.

The method we developed here is quite generally applicable to geological processes that have anything to do with mass balance or with any conserved quantities. One immediate application is a stochastic map of radiogenic heat production from U and Th, shown in Fig. 11 (numerical data file is also provided in the supplementary material).

## 7 Conclusions

For quantitative estimation of local geoneutrino fluxes, we developed a new fully-probabilistic framework. A 3-D lithological map was constructed by Bayesian inference of lithology from seismic tomography observation, using a prior model derived from a local geological exposure. Despite the fact that the prior does not include any vertical information, the inferred lithology model revealed a vertical structure that is consistent with our general picture of the continental crust. Unlike the conventional picture, however, our output is reproducible and has uncertainty estimations.

For the compositional models, we emphasize the importance of removing all possible biases. We showed that the commonly-used log normal model produces a bias in sum and mean values, and we introduced an alternative gamma distribution model that conserves those values if the model is fit to samples using the maximum likelihood method. The gamma model can fit both highly-skewed distributions, which are common for igneous rocks, as well as Gaussian-like distributions, which are common for sedimentary rocks. This universality might not be just by chance, as the reproductive property of the gamma distribution is also present in rock mixing processes. We also developed a method to properly treat the samples below detection limits; this technique can also be used to eliminate analysis instabilities due to rare samples in distribution tails. Other biases remain, such as bias in researchers' interest on collecting samples. All biases must be eliminated eventually, but in this work we focused on removing biases in our methodology framework with identifying a number of biases existing in past treatments.

By combining the lithology model PDF and composition model PDF, the geoneutrino flux PDF was calculated, representing the first construction of a fully probabilistic flux model.

Currently the uncertainties are large at ~70% because of the difficulty in handling correlations. Experience suggests that the majority of the correlation effects might have been washed away by averaging processes in integration. An investigation into this with more geological data is ongoing.

Owing to the mechanism to incorporate localities, the method presented here is immediately applicable to geoneutrino modeling of other regions. It is also applicable to global modeling by inserting into the framework a proper lithology prior and rock compositions.

## Acknowledgments

We thank W. F. McDonough, F. Mantovani and J. Detwiler for discussions and comments and K. Ozawa for providing some of the xenolith samples analyzed in this study. This work was supported by JSPS KAKENHI Grant Number 15H05833. The tomography model used in this study is available on the web page of NIED JAPAN. All the other input and output data of this study are explicitly presented in the tables or supplementary information.

# Appendix

## A1. Methods for Lithology Identification for Two Component System

The lithology PDF model discussed in Section 3 assumes single lithology in each cell. Here we show that this method can be naturally extended for mixed lithology, with an example of two lithology mixture. For the two component system, we infer the conditional probability density function for the phenomenon that $i$-th and $j$-th lithology are mixed by the ratio of $(1-q){:}\,q$ $(0 < q < 1/2)$, $P(ij, q|v^{obs})$, where both of $i$-th and $j$-th rock-types are elements in the rock-type-group $S^p$ that is a set of rock-types appearing in our prior (i.e., $i, j \in S^p$ and $i \neq j$). By definition we have

$$\sum_{i,j\ (i\neq j)} \int_0^{1/2} P(ij, q|v^{obs})\, dq = 1.$$

Using the laboratory experiment data, we can infer the expected value and variance of seismic velocity of that mixed lithology as

$$v^{(ij,q)} = (1-q)v^{(i)} + qv^{(j)} \text{ and}$$

$$\sigma^{(ij,q)^2} = (1-q)\sigma^{(i)^2} + q\,\sigma^{(j)^2}.$$

In the similar fashion, we infer the temperature coefficient of the mixed lithology as

$$\alpha^{(ij,q)} = (1-q)\alpha^{(i)} + q\,\alpha^{(j)}.$$

Using these, we have

$$P(v^{obs}|ij, q) = \frac{1}{\sqrt{2\pi}\,\sigma^{all}} \exp\left(-\frac{\left(v^{obs} - v^{(ij,q)}\right)^2}{2\sigma^{all^2}}\right),$$

where $\sigma^{all^2} = \sigma^{obs^2} + \alpha^{(ij,q)^2}\sigma^{temp^2} + \sigma^{(ij,q)^2}$. We can then evaluate the conditional probability density function $P(ij, q|v^{obs})$ by

$$P(ij, q|v^{obs}) = \frac{P(v^{obs}|ij, q)\, P(ij, q)}{\sum_{i',j'(i'\neq j')} \int P(v^{obs}|i'j', q)\, P(i'j', q)\, dq},$$

where $P(ij, q)$ is the prior. We start from a simple model that the prior is proportional to internally dividing quantity of $P(i)$ and $P(j)$:

$$P(ij, q) \propto (1-q)P(i) + qP(j),$$

or

$$P(ij, q) = \frac{(1-q)P(i) + qP(j)}{\sum_{i',j'(i'\neq j')} \int [(1-q)P(i') + qP(j')]dq}.$$

In the future, we can include more information into the prior such as gravitational observations and/or chemical potentials. The case without mixture corresponds to the case when we fix $q = 0$. Using this conditional probability, we evaluate the marginal probability $P(i|v^{obs})$ by

$$P(i|v^{obs}) = \begin{cases} P(\, 5.0 \le v < 7.5 \mid v^{obs})\; P^0(i|v^{obs}) & (1 \le i \le 29) \\ P(\, v < 5.0 \mid v^{obs}) & (i = 30) \\ P(\, 7.5 \le v \mid v^{obs}) & (i = 31) \end{cases},$$

where

$$P^0(i|v^{obs}) = \sum_{i,j\ (i \ne j)} \int (1-q)\, P(ij, q|v^{obs})\ dq + \sum_{j,i\ (j \ne i)} \int q\, P(ji, q|v^{obs})\ dq.$$

## A2. Chemical analysis of xenoliths from the Japan arc

In this study, we have determined major and trace element abundances in 203 xenoliths from the Japan arc (48 samples from Ichinomegata, 91 samples from Oki-dogo, and 64 samples from Takashima; see Fig. 6 for the localities). To minimize the effect of contamination by host magma, we analyzed only parts of xenolith samples >1 cm away from the host rocks. A few gram pieces of xenolith samples were roughly crushed with a metal disk mill, and then powdered with an agate mortar and pestle. The sample powders were fused in $Li_2B_4O_7$ (sample : $Li_2B_4O_7$ = 1 : 10) to form glass discs on which major and trace element analyses were conducted at the University of Tokyo Department of Earth and Planetary Science. The major element analysis was carried out using a Panalytical Axios X-ray fluorescence spectrometer. The operating conditions were 60 kV accelerating voltage and 66 mA beam current for Fe and Mn, 40 kV and 100 mA for Ti, and 32 kV and 125 mA for Na, Mg, Al, Si, P, K, and Ca. The X-rays were diffracted by LiF, Ge, PE, and PX1 crystals and measured by a gas flow detector. The trace element analysis was performed on a CETAC LSX-213 G2+ Nd:YAG laser ablation (LA) system attached to a Thermo Fisher Scientific icap Qc inductively coupled plasma mass spectrometer (ICPMS). The details of the instruments are given by Itano et al. (2016). The data were acquired over a period of ~5 min under LA sampling conditions with a spot size of 100 µm, a rastering rate of 5 µm/s, and a repetition rate of 4 Hz. Helium gas was utilized for flushing the ablation site to minimize aerosol deposition. The Ca concentrations measured by XRF were used as the internal standard, and LA-ICPMS sensitivity factors for U and Th relative to Ca were determined from the results of NIST SRM 613 glass analyses (Pearce et al., 1997).

## A3. Maximum likelihood fitting of Gamma Distribution

Here we show that if a gamma distribution PDF is fit to samples using the maximum likelihood method, the expected value of the fit PDF is always equal to the mean value of the samples.

Using a shape parameter $k > 0$ and a scale parameter $\theta > 0$, the PDF of gamma distribution is

$$f(x; k, \theta) = \frac{x^{k-1} e^{-x/\theta}}{\Gamma(k)\theta^k}$$

and its expected value is given by

$$E\{x\} = k\theta$$

For the samples $\{a_1, a_2, \ldots, a_N\}$, the log-likelihood is

$$\log \mathcal{L}(k,\theta) = \sum_{j=1}^{N} \log\left[f\left(a_j; k, \theta\right)\right]$$

$$= (k-1)\sum_{j=1}^{N} \log a_i - \frac{1}{\theta}\sum_{j=1}^{N} a_j - N[\Gamma(k) + k \ \log\theta]$$

To find the maximum with respect to $\theta$, take the derivative

$$\frac{\partial \log \mathcal{L}}{\partial \theta} = \frac{1}{\theta^2}\sum_{j=1}^{N} a_j - Nk\frac{1}{\theta}$$

The maximum likelihood estimators (MLE) of $\theta$ and $k$, $\hat{\theta}$ and $\hat{k}$, are given by the values that make this derivative (as well as $\frac{\partial \log \mathcal{L}}{\partial \theta}$) equal to zero. This gives,

$$\hat{k}\,\hat{\theta} = \frac{1}{N}\sum_{j=1}^{N} a_j$$

Therefore the best-fit gamma-distribution PDF, $f\left(x; \hat{k}, \hat{\theta}\right)$, has the expected value equal to the mean value of the samples:

$$\hat{E}\{x\} = \hat{k}\,\hat{\theta} = \frac{1}{N}\sum_{j=1}^{N} a_i$$

This shows that fitting to a gamma model preserves the sample mean, for any underlying distribution.

Next we show that the same is not true for the log-normal distribution. For the log-normal distribution,

$$\mathrm{LogNormal}(x; \mu, \sigma^2) = \frac{1}{\sqrt{2\pi\sigma^2}x} e^{-\frac{(\log x - \mu)^2}{2\sigma^2}}$$

the expected value is given by

$$E\{x\} = \exp\left[\mu + \frac{\sigma^2}{2}\right]$$

The MLEs of $\mu$ and $\sigma^2$, $\hat{\mu}$ and $\widehat{\sigma^2}$, are given by

$$\hat{\mu} = \frac{1}{N}\sum_{j=1}^{N} \log(a_i) \quad \text{and} \quad \widehat{\sigma^2} = \frac{1}{N}\sum_{j=1}^{N} [\log(a_i) - \hat{\mu}]^2 \ .$$

It is difficult to derive the relationship between $\hat{E}\{x\}$ and the sample mean in this case, however for our purpose it is sufficient to show that they are not equal with a simple example. Consider the case that the samples are two rocks with compositions of 1 and $e$ (in arbitrary unit). The MLEs are calculated to be $\hat{\mu} = 1/2$ and $\widehat{\sigma^2} = 1/8$, yielding the expected value of the fit distribution to be $e^{9/16} \sim 1.75$, which is different from the sample mean, $(1+e)/2$ ~ 1.86. Also note that the median value of the fit PDF (equivalent to calculating the sample geometric mean), $\exp(\hat{\mu}) \sim 1.65$, is different from the sample mean.

## A4. Sampling Bias

Our aim in this work is to eliminate bias in the *method* used to calculate the geoneutrino flux from input seismological and geochemical datasets. Biases in input dataset themselves are beyond our scope. However, since the biases in the input datasets will still be propagated to the final result, we evaluate the impact of them.

Biases in sample collection can be classified into two cases; the case that the rock is too rare, and the case that the rock is too common. If rocks are too rare, the representativeness of collected samples could be questionable, because there might be a specific reason why the rare rock samples are brought or exposed to the accessible location. However, such rocks will not largely affect our flux calculation because those are either low in total mass and not contributing to the total flux, or are deep-origin rocks that are low in U and Th in general.

If rocks are too common, sample collection might be biased by collectors' interest. Some might seek for typical and/or clean samples, while some might be more interested in samples with unique features. Since commonplace rocks are shallow and therefore enriched in U and Th in general, this will be a major source of bias in our flux calculation. There might be another bias at the analysis stage, namely selecting samples for final analysis based on rough pre-analysis to assure coverage, but this will only make the distribution wider leading to a more conservative result.

Rocks that are problematic for being too common are typically found near the surface of the crust. We evaluate the effect of the bias by comparing the surface part of our inferred model with two previous studies based on systematic surface rock sampling. Both of these studies aim for a bias-free estimation of the crustal composition of Japan. Fig. A9 in Appendix shows the PDFs for average composition estimations of Japan Arc surface, calculated from our model assuming the maximum correlation, from Togashi et al. (2000) and from Japan Geochemical Map (Imai et al., 2004). These publications do not provide error estimations. To construct PDFs in the figure, bootstrapping is used. For additional comparison, the global average crustal model by Rudnick and Gao (2003) is also shown. Since these estimations are made using different approaches (outcrop samples vs river sediment), some differences are expected in addition to the sampling bias by interest. Given that significant differences are apparent, we cannot exclude the existence of sampling bias in some or all of the estimations, but the spread is smaller than our current error range and therefore such sampling bias will not significantly affect our current result. If our maximum correlation assumption could be eliminated in the future, we will have to solve the sampling bias problem. Joint investigations with the authors of these articles is under way.

**Table 1.** The proportion of rock types exposed in the Hidaka metamorphic belt. The categorization of lithology follows that by Christensen and Mooney (1995). The original classifications in Shimura et al. (2004) are also presented for reference.

| Christensen & Mooney | Shimura et al. | Percentage |
|---|---|---|
| Metagraywacke (MGW) | unmetamorphosed sediment | 8.6% |
| Slate (SLT) | low-grade metasediment | 6.5% |
| Phyllite (PHY) | semi-schist, pelitic-schist | 10.0% |
| Biotite (Tonalite) Gneiss (BGN) | middle, basal, and lower tonalite | 18.6% |
| Granite-Granodiorite (GRA) | upper granite | 3.4% |
| Mica Quartz Schist (QSC) | gneisses, high-grade gneisses | 7.8% |
| Paragranulite (PGR) | pelitic granulite | 10.1% |
| Mafic Granulite (MGR) | mafic granulite | 7.4% |
| Amphibolite (AMP) | amphibolite | 6.7% |
| Gabbro-Norite-Troctolite (GAB) | upper gabbro, gabbro | 16.9% |
| Mafic Garnet Granulite (GGR) | restitic granulite | 4.0% |

**Table 2.** Criteria for geochemical rock classification.

| Rock type and abbreviation in geophysical modeling | Rock type in geochemical modeling | Description |
|---|---|---|
| Amphibolite (AMP) | AMP | Holocrystalline rocks with mineral assemblage corresponding to amphibolite facies. |
| Biotite (Tonalite) gneiss (BGN) | BGN | Holocrystalline rocks with mineral assemblage of tonalite, including gneissic rock. |
| Gabbro-Norite-Troctolite (GAB), Mafic granulite (MGR) and Mafic garnet granulite (GGR) | GAB/MGR/GGR | Mafic holocrystalline rocks with mineral assemblage of gabbro or granulite. Garnet-absent granulite and Garnet-bearing granulite were grouped together. No systematic differences in U and Th concentrations are observed in our database, possibly because U and Th are incompatible both with mineral assemblages of pyroxene granulite and of garnet granulite. GAB/MGR/GGR were grouped according to their $SiO_2$ concentrations following the International Union of Geological Sciences (IUGS) classification of igneous rocks (Le Bas & Streckeisen, 1991). Gabbro and granulite with $SiO_2$ < 52 wt% are classified into this group. |
| Granite-Granodiorite (GRA) and Granite Gneiss (GGN) | GRA/GGN | Holocrystalline rocks with mineral assemblages of granite or granodiorite, including gneissic rock. Aplite and small-scale dikes were not used. |
| Metagraywacke (MGW), Slate (SLT) and Phyllite (PHY) | MGW/SLT/PHY | Pelitic origin, fine-grained, foliated weakly metamorphosed rock. U and Th concentrations of MGW/SLT/PHY are also used to model the composition of surface sediment (SED). |
| Mica quartz schist (QSC) | QSC | Sediment origin (micro- or geological structure of original sedimentary rock, or cordierite or kyanite-bearing) holocrystalline rocks with schistosity structure. |
| Paragranulite (PGR) | PGR | Sediment origin (micro- or geological structure of original sedimentary rock, or cordierite or kyanite-bearing) holocrystalline rocks with mineral assemblage of granulite facies. |
| Mantle | MTL | Compositions of mantle peridotite xenolith compiled using the GEOROC database are used. |
| Surface sediment | SED | U and Th concentrations of MGW/SLT/PHY are used. |

**Table 3.** Sample statistics.

| Rock type | Element (ppm) | Mean | Median | n | Mean (X≥0.2) | Median (X≥0.2) | n (X≥0.2) |
|---|---|---|---|---|---|---|---|
| AMP | U | 0.45 | 0.32 | 50 | 0.58 | 0.40 | 37 |
| | Th | 3.37 | 1.16 | 61 | 4.17 | 1.85 | 49 |
| BGN | U | 1.31 | 1.18 | 63 | 1.32 | 1.19 | 62 |
| | Th | 4.55 | 4.00 | 47 | | | |
| GAB/MGR/GGR | U | 0.16 | 0.09 | 218 | 0.47 | 0.32 | 50 |
| | Th | 0.35 | 0.15 | 216 | 0.71 | 0.43 | 93 |
| GRA/GGN | U | 3.12 | 2.52 | 203 | | | |
| | Th | 13.57 | 11.70 | 198 | | | |
| MGW/SLT/PHY | U | 2.38 | 2.41 | 71 | | | |
| | Th | 12.44 | 13.30 | 71 | | | |
| MTL | U | 0.12 | 0.04 | 747 | 0.62 | 0.41 | 96 |
| | Th | 0.26 | 0.09 | 763 | 0.78 | 0.40 | 203 |
| PGR | U | 1.74 | 1.68 | 12 | | | |
| | Th | 7.33 | 7.30 | 12 | | | |
| QSC | U | 2.17 | 2.24 | 19 | | | |
| | Th | 10.53 | 10.35 | 13 | | | |

**Table 4.** Parameter values of PDFs obtained by maximum likelihood estimation.

| | | Gamma | | | | | | Normal | | | | Lognormal | | | | | |
|---|---|---|---|---|---|---|---|---|---|---|---|---|---|---|---|---|---|
| Rock type | Element | *k* | *θ* | Mean | Median | Mean (X≥0.2) | Median (X≥0.2) | Mean | SD | Mean (X≥0.2) | Median (X≥0.2) | *μ* | *σ* | Mean | Median | Mean (X≥0.2) | Median (X≥0.2) |
| AMP | U | 1.38 | 0.33 | 0.46 | 0.35 | 0.60 | 0.49 | 0.46 | 0.42 | 0.65 | 0.60 | -1.10 | 0.81 | 0.46 | 0.33 | 0.58 | 0.44 |
| | Th | 0.49 | 6.92 | 3.36 | 1.51 | 4.21 | 2.36 | 3.37 | 5.16 | 5.70 | 5.14 | 0.13 | 1.70 | 4.86 | 1.14 | 5.69 | 1.58 |
| BGN | U | 2.31 | 0.57 | 1.31 | 1.13 | 1.35 | 1.16 | 1.31 | 0.85 | 1.47 | 1.41 | 0.04 | 0.72 | 1.35 | 1.04 | 1.36 | 1.05 |
| | Th | 3.20 | 1.42 | 4.55 | 4.09 | | | 4.55 | 2.49 | | | 1.35 | 0.61 | 4.66 | 3.86 | | |
| GAB/MGR/GGR | U | 0.36 | 0.40 | 0.14 | 0.05 | 0.48 | 0.38 | 0.19 | 0.25 | 0.40 | 0.36 | -2.47 | 1.15 | 0.16 | 0.08 | 0.47 | 0.34 |
| | Th | 0.43 | 0.79 | 0.34 | 0.13 | 0.73 | 0.53 | 0.37 | 0.61 | 0.75 | 0.68 | -1.83 | 1.27 | 0.36 | 0.16 | 0.72 | 0.44 |
| GRA/GGN | U | 2.41 | 1.30 | 3.12 | 2.70 | | | 3.12 | 2.32 | | | 0.92 | 0.67 | 3.13 | 2.50 | | |
| | Th | 3.54 | 3.83 | 13.56 | 12.31 | | | 13.56 | 7.79 | | | 2.46 | 0.55 | 13.61 | 11.70 | | |
| MGW/SLT/PHY | U | 13.58 | 0.18 | 2.38 | 2.32 | | | 2.38 | 0.59 | | | 0.83 | 0.30 | 2.40 | 2.30 | | |
| | Th | 8.80 | 1.41 | 12.44 | 11.97 | | | 12.44 | 3.28 | | | 2.46 | 0.40 | 12.72 | 11.74 | | |
| MTL | U | 0.12 | 0.78 | 0.10 | $1.49\times10^{-3}$ | 0.62 | 0.45 | 0.16 | 0.33 | 0.45 | 0.41 | -3.41 | 1.58 | 0.12 | 0.03 | 0.61 | 0.37 |
| | Th | 0.21 | 1.09 | 0.23 | 0.03 | 0.79 | 0.54 | 0.27 | 0.50 | 0.63 | 0.57 | -2.59 | 1.57 | 0.26 | 0.08 | 0.80 | 0.43 |
| PGR | U | 5.60 | 0.31 | 1.74 | 1.64 | | | 1.74 | 0.74 | | | 0.46 | 0.44 | 1.75 | 1.59 | | |
| | Th | 2.36 | 3.10 | 7.33 | 6.33 | | | 7.33 | 3.74 | | | 1.77 | 0.83 | 8.23 | 5.84 | | |
| QSC | U | 21.12 | 0.10 | 2.17 | 2.14 | | | 2.17 | 0.46 | | | 0.75 | 0.22 | 2.18 | 2.12 | | |
| | Th | 10.11 | 1.04 | 10.53 | 10.18 | | | 10.53 | 3.17 | | | 2.30 | 0.33 | 10.57 | 10.01 | | |

**Table 5.** Log-likelihood values ($-2\log\mathcal{L}$) from maximum likelihood estimation.

| Rock type | Element | Gamma | Normal | Log-normal |
|---|---|---|---|---|
| AMP | U | 7.83 | 43.73 | 5.83 |
| | Th | 231.29 | 355.48 | 236.82 |
| BGN | U | 134.86 | 153.07 | 138.23 |
| | Th | 210.70 | 219.14 | 214.50 |
| GAB/MGR/GGR | U | -28.91 | 327.57 | -33.52 |
| | Th | 53.74 | 497.42 | 43.75 |
| GRA/GGN | U | 797.72 | 918.18 | 784.62 |
| | Th | 1304.02 | 1374.56 | 1299.10 |
| MGW/SLT/PHY | U | 26.52 | 126.57 | 146.34 |
| | Th | 399.57 | 370.09 | 421.02 |
| MTL | U | 6.52 | 2251.65 | -9.09 |
| | Th | 141.09 | 2518.97 | 116.48 |
| PGR | U | 25.15 | 26.79 | 25.66 |
| | Th | 67.78 | 65.71 | 71.90 |
| QSC | U | 24.85 | 24.21 | 25.60 |
| | Th | 67.15 | 66.91 | 67.88 |

**Table 6.** Comparison with previous studies. The errors of geoneutrino flux at Kamioka due to uncertainties of local upper crustal models are listed. Each study uses distinctive definition of local / near-field crust, therefore comparison of absolute flux cannot be made. To make the uncertainties comparable, some adjustments are applied as described in the table. Due to difficulties of the adjustment, only the upper continental crust parts are compared, where the local upper crust contributes ~3/4 of the flux from the entire local crust. Even for the upper continental crust, the adjustments are approximate as described in the table, therefore the comparison is not exact.

| | Fiorentini et al. (2005) | Enomoto (2005) ;<br>Enomoto et al. (2007) | Huang et al. (2013) [*1] | This work | This work (averaging assumed) |
|---|---|---|---|---|---|
| Flux error(s) due to crustal composition uncertainties | 10% | U: 14%<br>Th: 18% | U: 21%<br>Th: 10% | (U: 69%)<br>(Th: 61%) | (U: 35%)<br>(Th: 25%) |
| Flux error(s) due to crustal boundary uncertainties | 6% | 6% | (12%) | | |
| Error estimation methods | The surface exposed geology is assumed to extend vertically in the entire upper crust.<br><br>Composition error is the accuracy of chemical analysis of rock samples (3~4%), inflated by a factor of 3.<br><br>Crustal boundary error is from the “1 km accuracy for determining Moho and Conrad discontinuities”, inflated by a factor of 3. | The surface exposed geology is assumed to extend vertically to a depth of 15 km. The rest is assumed to be uniform with the composition of a global average model.<br><br>Composition errors are estimated by comparing the constructed local crustal model with a global model, and by comparing several rock composition models.<br><br>Crustal boundary error is estimated by comparing two published crustal models (local and global). | The entire upper crust is assumed to be uniform with the composition of a global average model.<br><br>Composition errors are directly taken from the global model, which are errors of the global mean estimation (as opposed to dispersion of composition).<br><br>Crustal boundary error is estimated by comparing three published global crustal models. | Bayesian inference of 3-D lithology using seismic tomography, combined with probabilistic rock composition models.<br><br>Use of tomography eliminates crustal boundary modeling.<br><br>Maximum correlation among all values is assumed. | Bayesian inference of 3-D lithology using seismic tomography, combined with probabilistic rock composition models.<br><br>Use of tomography eliminates crustal boundary modeling.<br><br>For rock composition uncertainties (not crustal composition uncertainties), “error of mean estimation” is used. For the |
| | | | | | others, maximum correlation is assumed. |
| Adjustments for this comparison | | The paper gives the errors in the total flux originating from the local model uncertainties. The paper states ~50% of the total flux comes within 500 km radius. The factor 2 (=1/0.5) is used to obtain the errors of the local contribution. | The error due to crustal thickness uncertainty quoted here is the uncertainty of global crustal thickness determination. This value is basically consistent with the value calculated by subtracting the composition error from the Kamioka flux error, which should represent the error from the local crustal uncertainties. | The model does not have segmentation to upper, (middle) and lower crusts, therefore the error of the entire “local” region is quoted. The “upper crust only” error would be smaller than the values here. | The model does not have segmentation to upper, (middle) and lower crusts, therefore the error of the entire “local” region is quoted. The “upper crust only” error would be smaller than the values here. |

[*1)] Huang et al. (2013) provides the most comprehensive flux calculation to date with uncertainty estimations for various locations including Kamioka, but it does not primarily aim to investigate the uncertainties of the Kamioka-vicinity contributions discussed here.

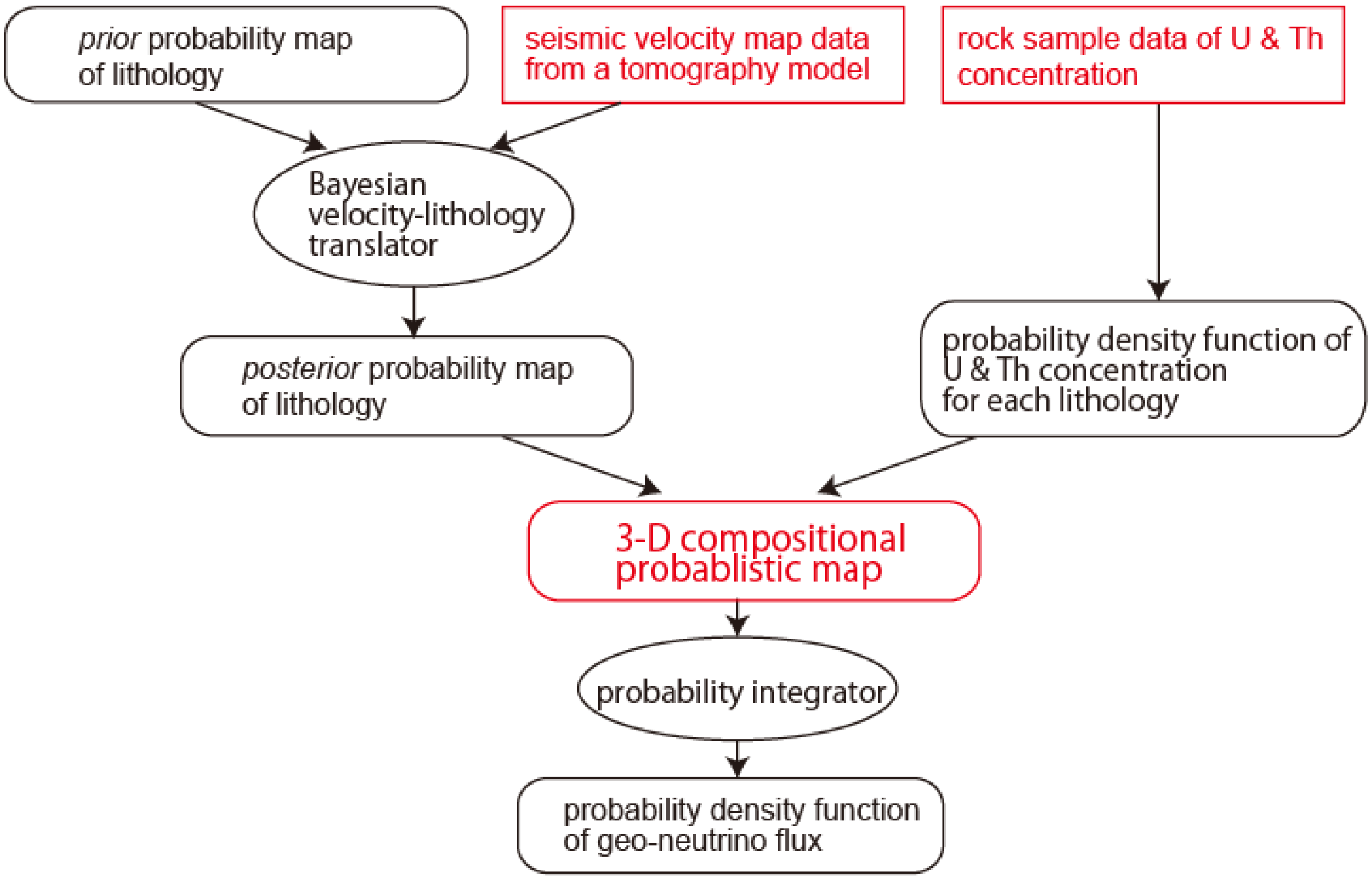

**Figure 1**. Flow chart of the neutrino flux calculation in this study. The rounded rectangles denote probability models, and the reds denote geological data.

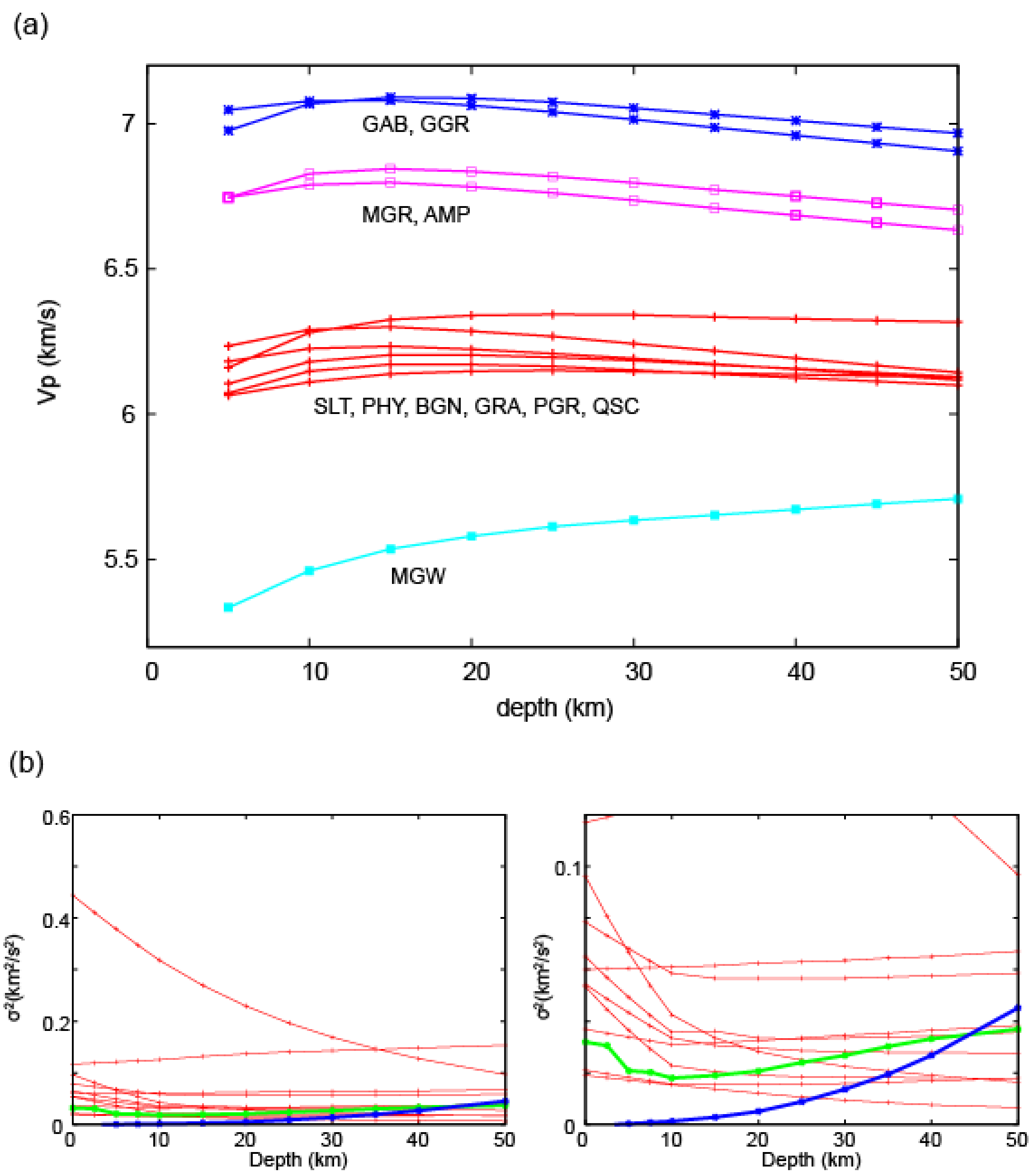

**Figure 2**. (a) Expected P velocities as a function of depth for the lithology-types appearing in our prior model. (b) Comparison of $\sigma^{(i)^2}$ (thin red lines), $\sigma^{obs^2}$ (thick green lines), and $\alpha^{(i)^2}\sigma^{temp^2}$ (thick blue lines) as a function of depth. The left and right panels are identical except the right is a blow-up of the left for smaller values of the vertical axis.

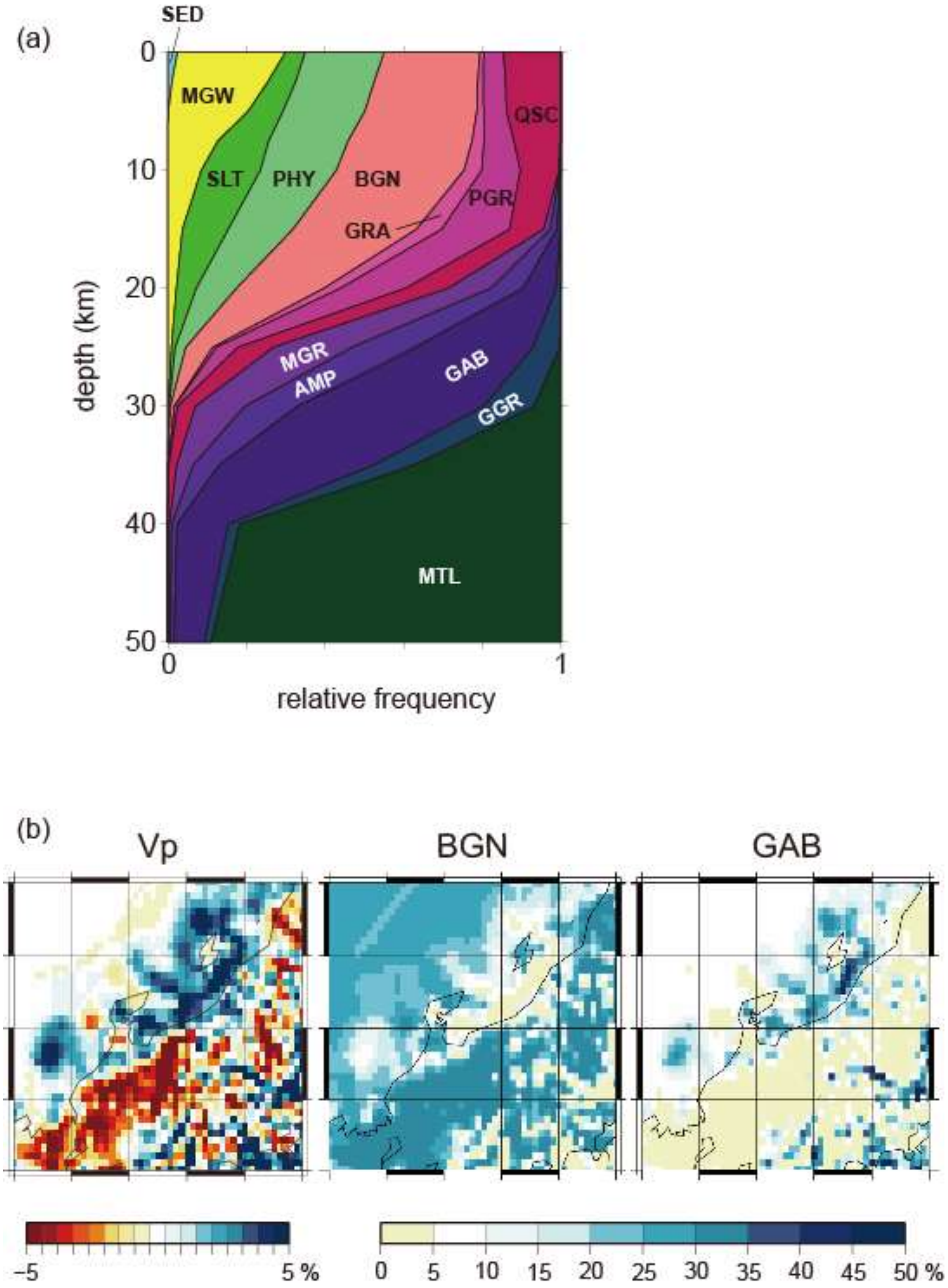

**Figure 3**. (a) Relative lithology frequencies at each depth obtained by averaging $P^{(x)}(i|v^{obs})$ over each horizontal plane. The resultant relative frequencies correspond to the expected bulk lithology at each depth. (b) Comparison of the tomography model we use (left), the posterior probability of BGN (center), and the posterior probability of GAB (right) at the 20 km depth.

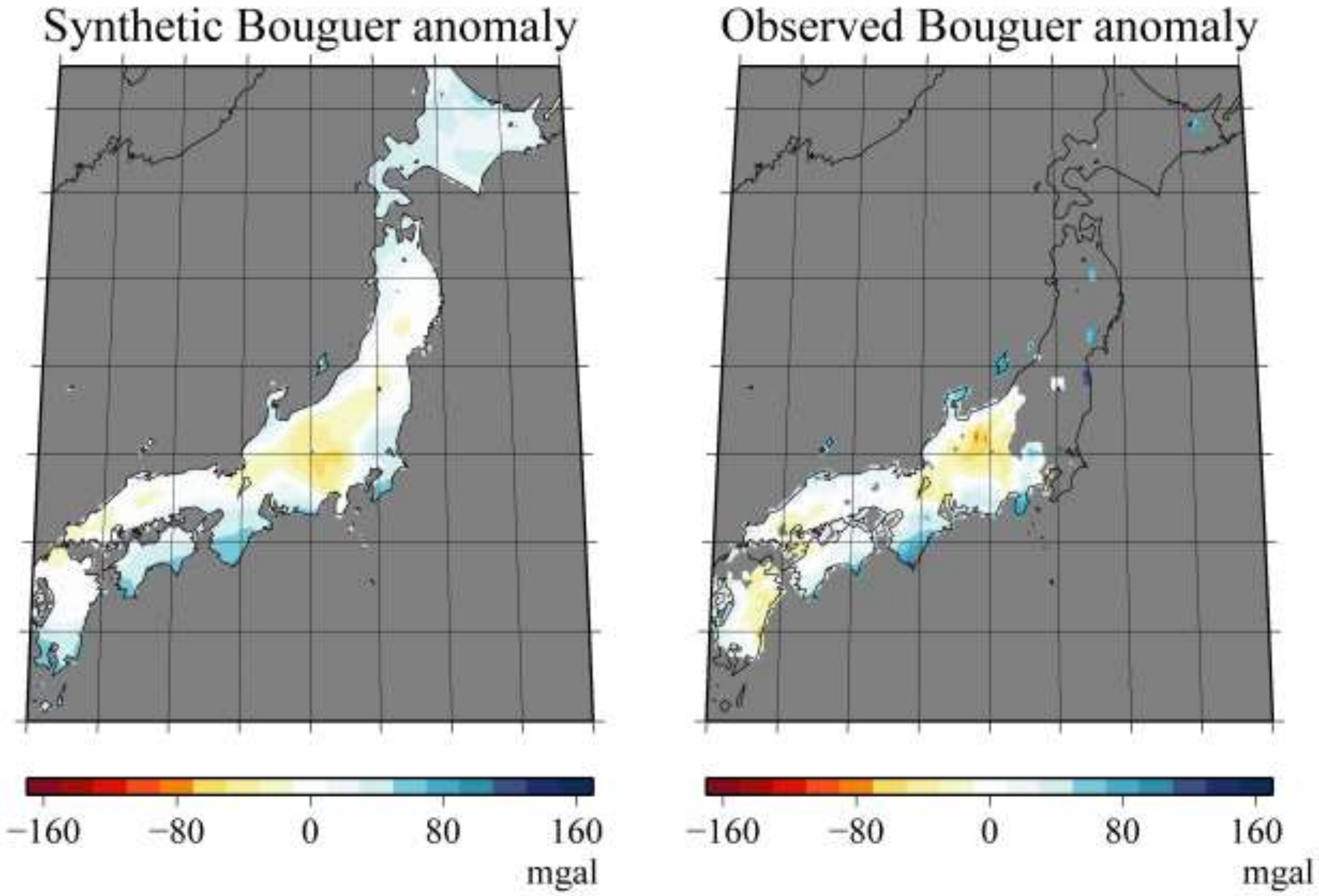

**Figure 4**. Comparison of the synthetic (left) and observed (right) Bouguer anomaly. The synthetic anomalies are computed for our probabilistic lithology model.

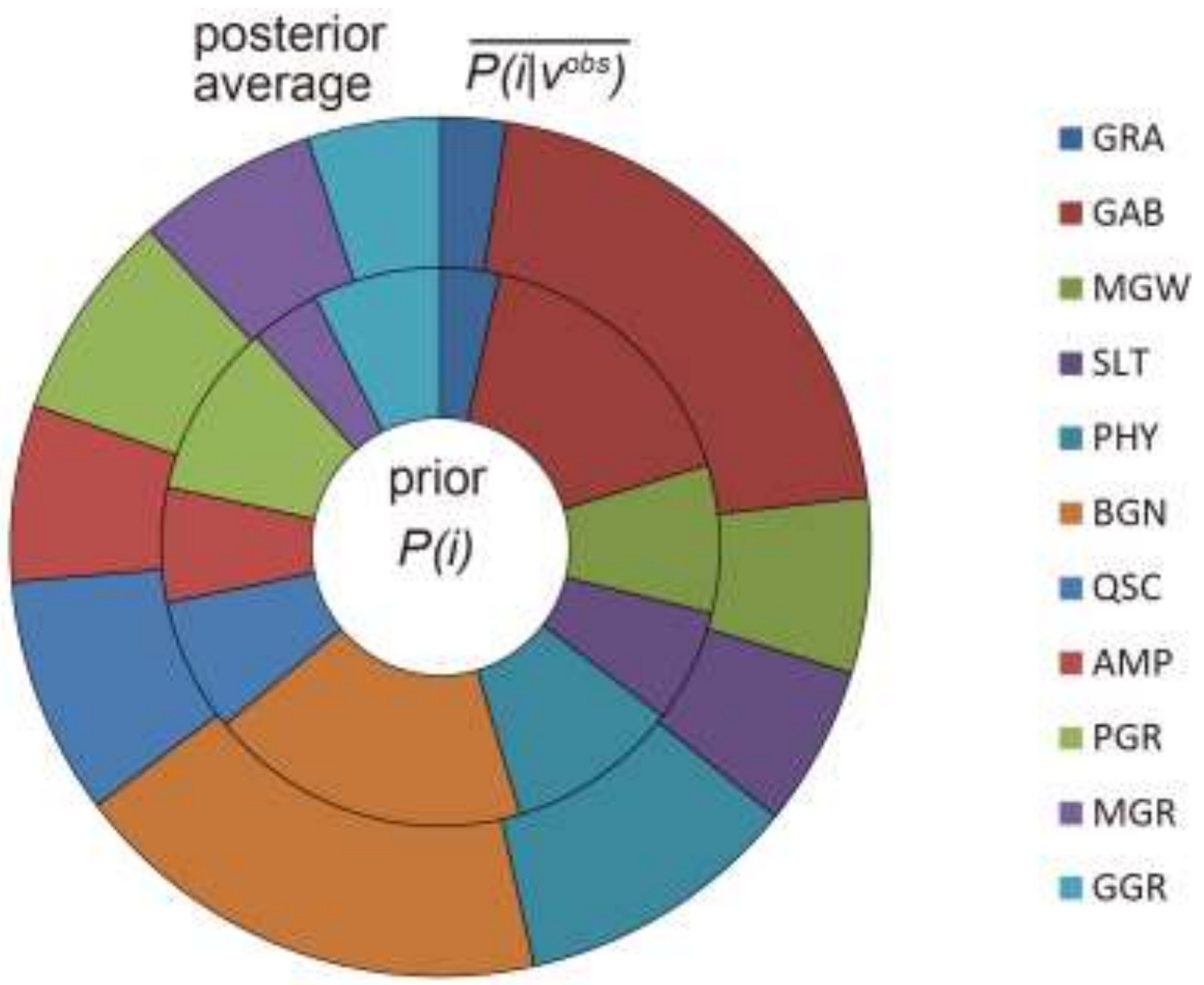

**Figure 5**. Comparison of the relative lithology frequencies in the prior and the posterior ensemble. Frequencies in the posterior ensemble are obtained by averaging $P^{(\boldsymbol{x})}(i|v^{obs})$ over the whole region. The resultant relative frequencies in the posterior ensemble correspond to the expected bulk lithology in the region of interest.

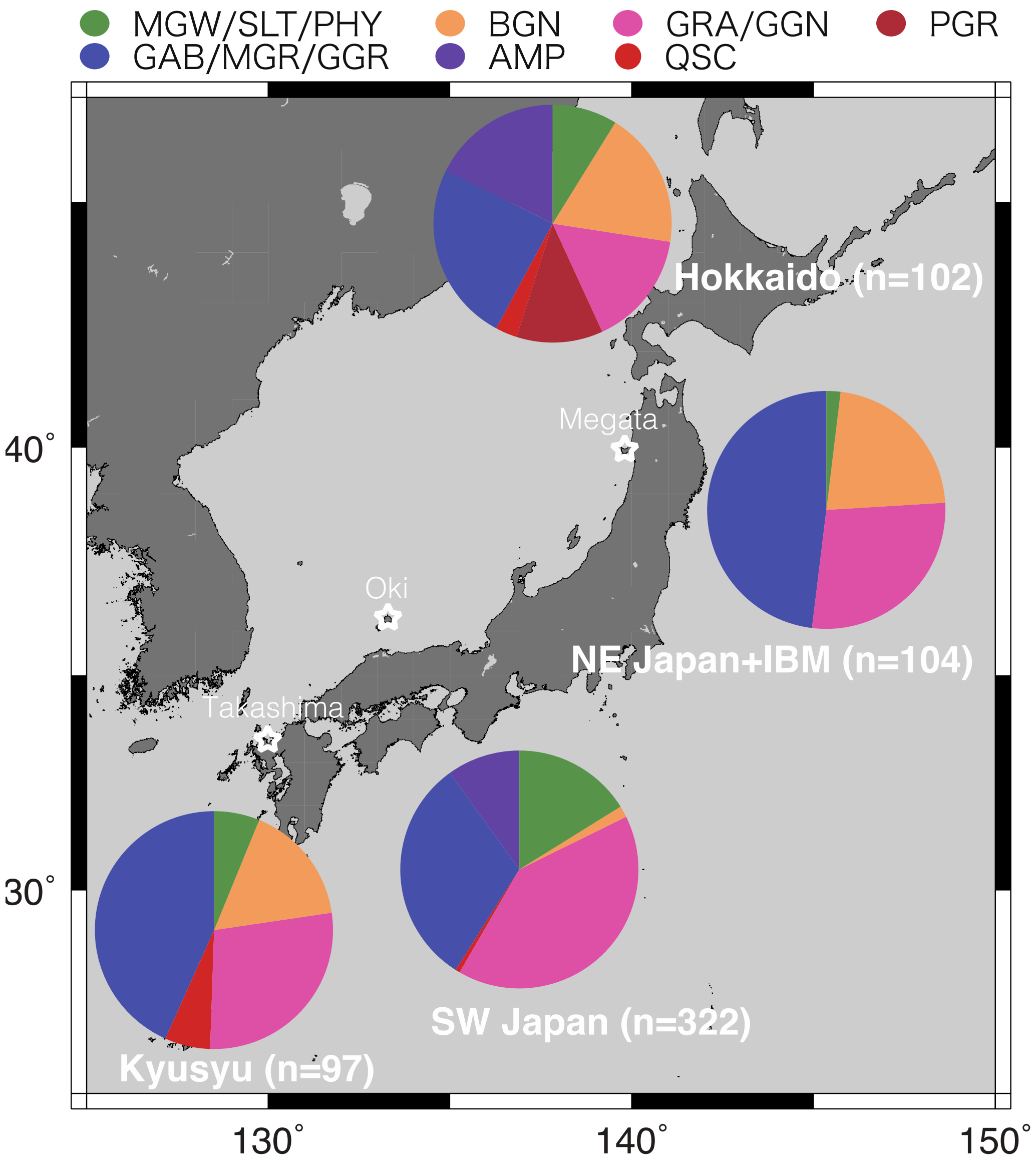

**Figure 6**. Numbers of samples, proportions of rock types, and locality map of xenolith samples. Proportions of rock types are shown by pie charts. Stars indicate localities of lower crustal xenoliths used in this study.

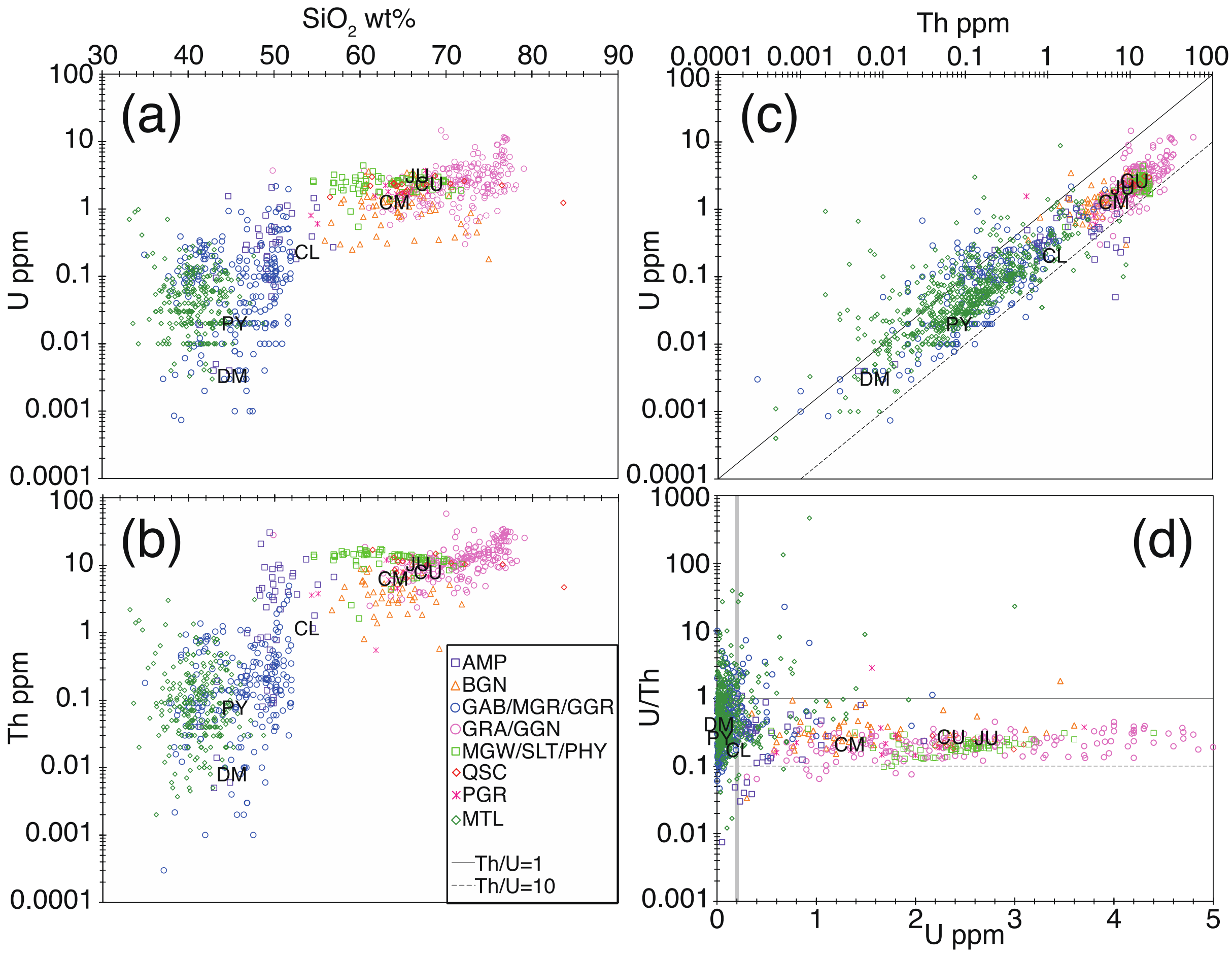

**Figure 7.** U, Th and $SiO_2$ concentrations of rocks used for the modeling. Black solid and dotted lines in Fig. 7c and 7d indicate U/Th=1 and U/Th=0.1, respectively. “JU” denotes average composition of Japan upper crust (Togashi et al., 2000), “CU”, “CM” and “CL” denote continental upper, middle and lower crust (Rudnick & Gao, 2003), “PY” denotes Pyrolite (McDonough & Sun, 1995) and “DM” denotes depleted MORB mantle (Workman & Hart, 2005). Gray vertical line in Fig. 7d indicate U=0.2 ppm (see text for detail).

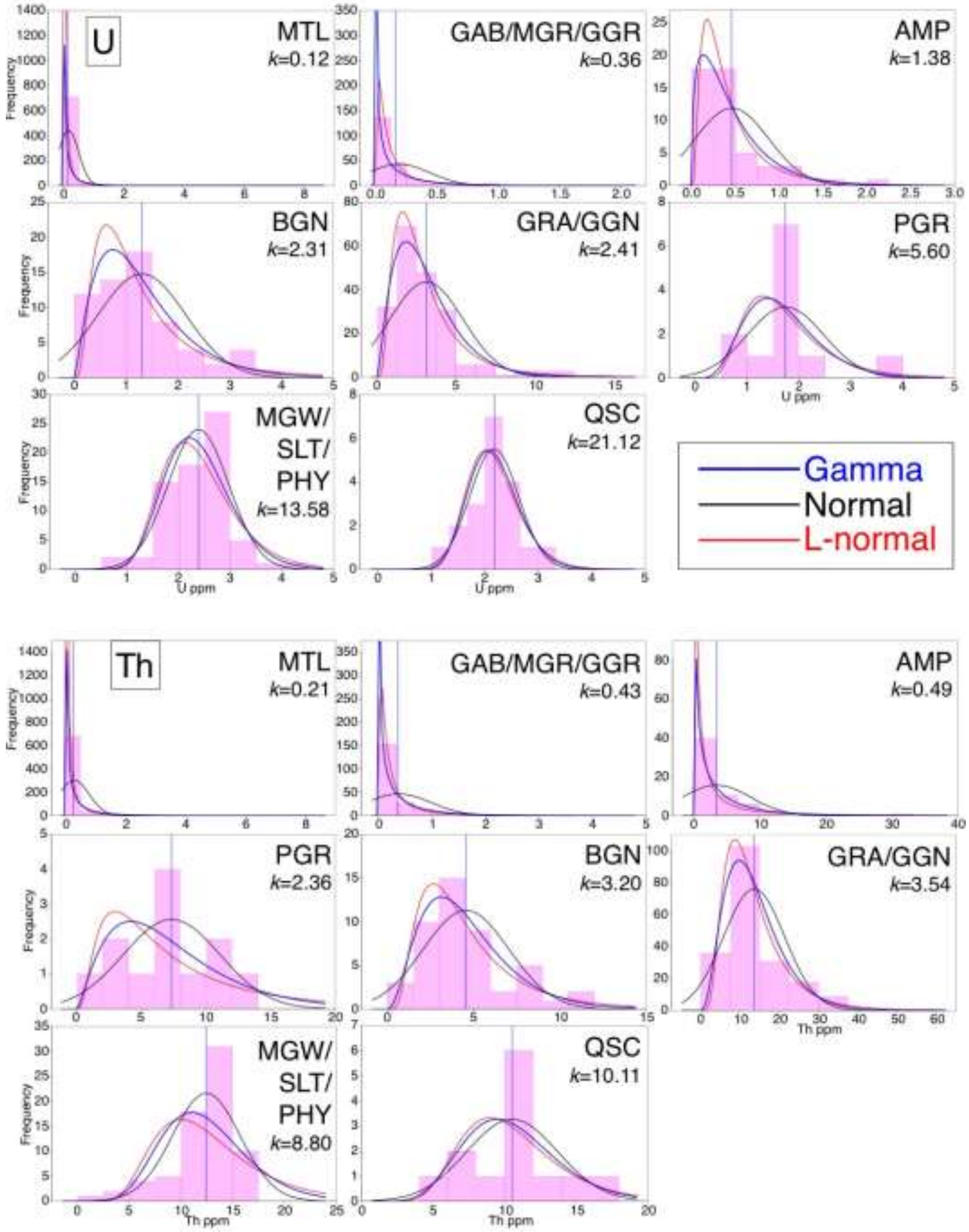

**Figure 8**. Histogrammed U and Th concentrations for different rock types and modeled probability density functions. Best-fit gamma, normal, and lognormal distributions are drawn as blue, black and red lines, respectively. The mean value of the data is drawn as a vertical blue line. See Section 4.3 for the *k* parameter.

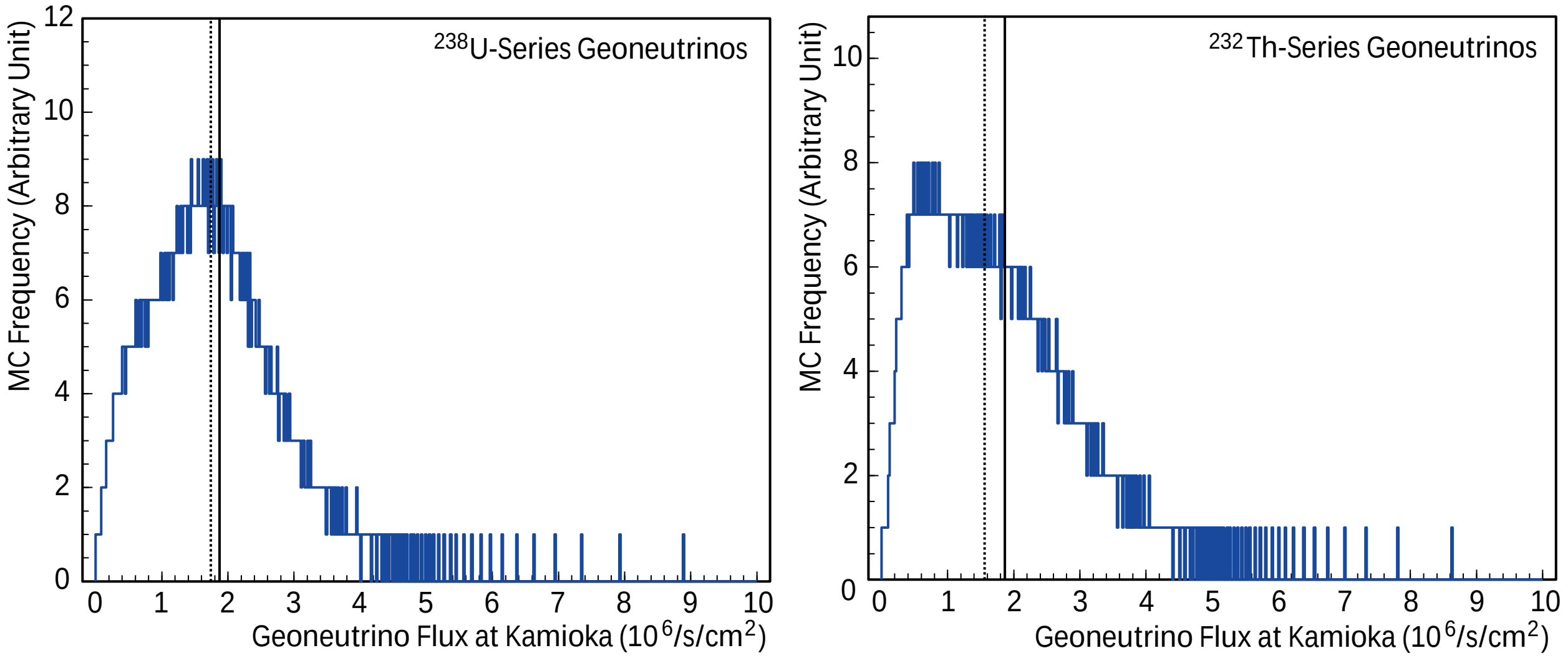

**Figure 9.** Geoneutrino flux PDF at Kamioka from $^{238}$U decays (left) and $^{232}$Th decays (right) in the Japan crust. The histograms are frequency distributions of fluxes from MC instances (implemented as "quantile scan" under the maximal correlation assumption; see text). The solid vertical lines indicate the mean flux values calculated from the mean U/Th concentration maps (where the mean fluxes are basically equivalent to the expected values of flux PDF), and the dotted vertical lines indicate the flux from previous models adjusted for the same geological region-of-interest.

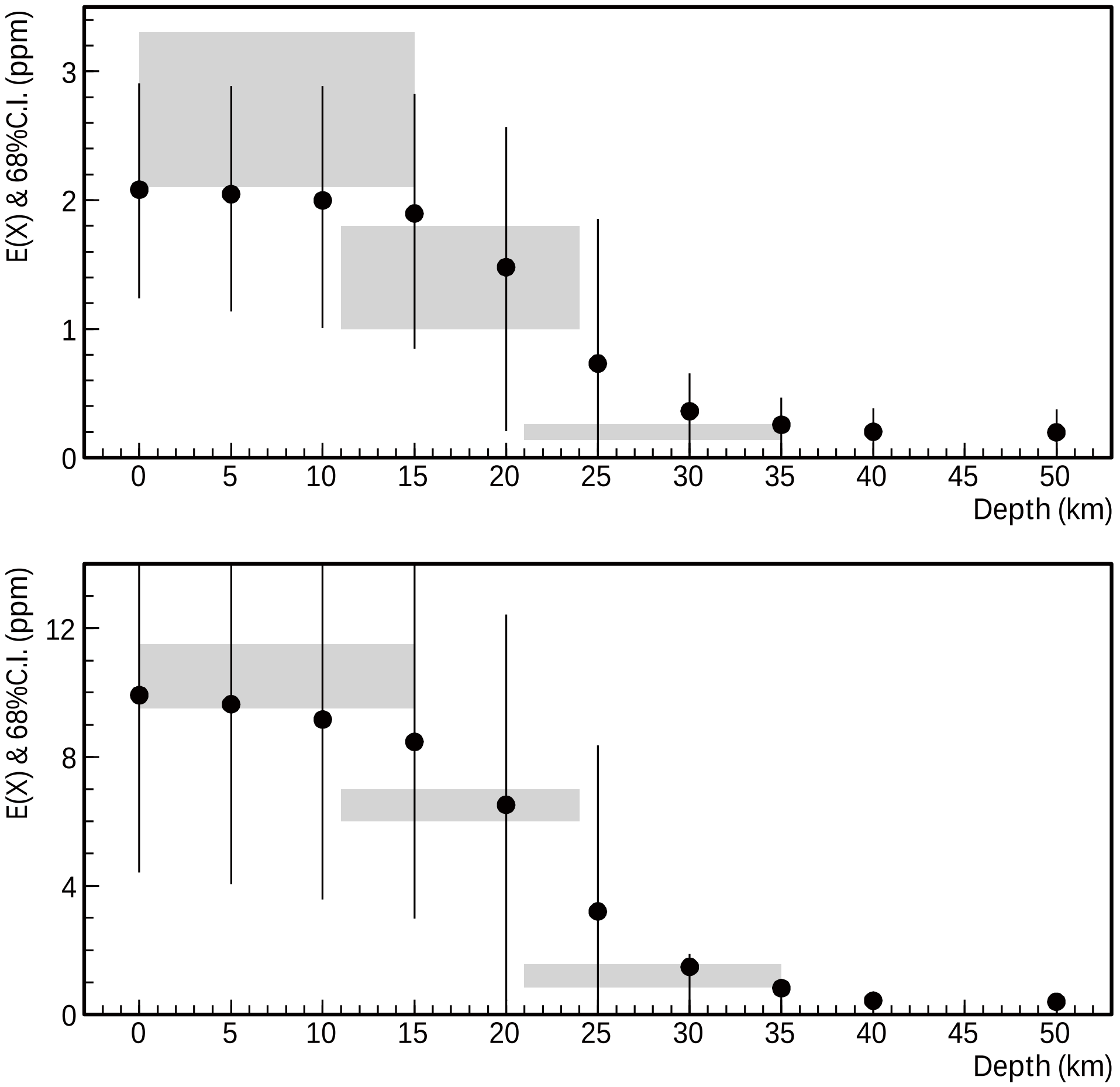

**Figure 10**. Comparison of Crustal models, for the vertical distributions of U (upper) and Th (lower) concentrations in the Japan crust. The points with error bars are calculated by combining our lithology model and our compositional model, assuming maximum correlations. The shared boxes are from CRUST 2.0 (horizontal span; from left to right, upper, middle and lower crust) and Rudnick and Gao (2003) (vertical span). Because the crustal boundaries are not at a constant depth, there are overlaps of crustal segmentations. Also note that this will smear the vertical compositional distributions.

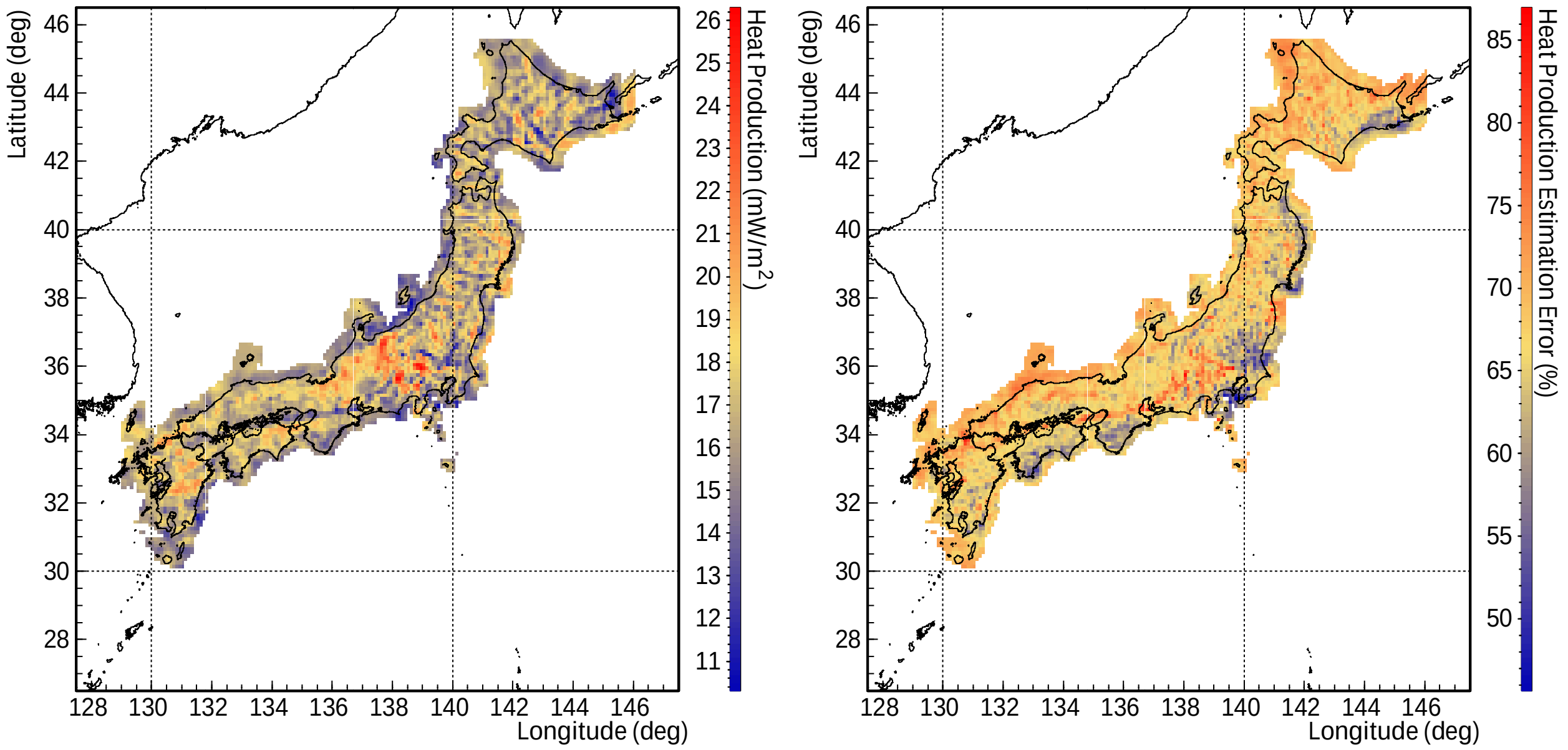

**Figure 11.** Left: radiogenic heat production map (vertical column total per unit area, U and Th combined), and right: with error estimation, constructed from the probabilistic lithology and composition models. For the error estimation, maximum correlation is assumed. A numerical table of the data shown here is provided in the supplementary material.

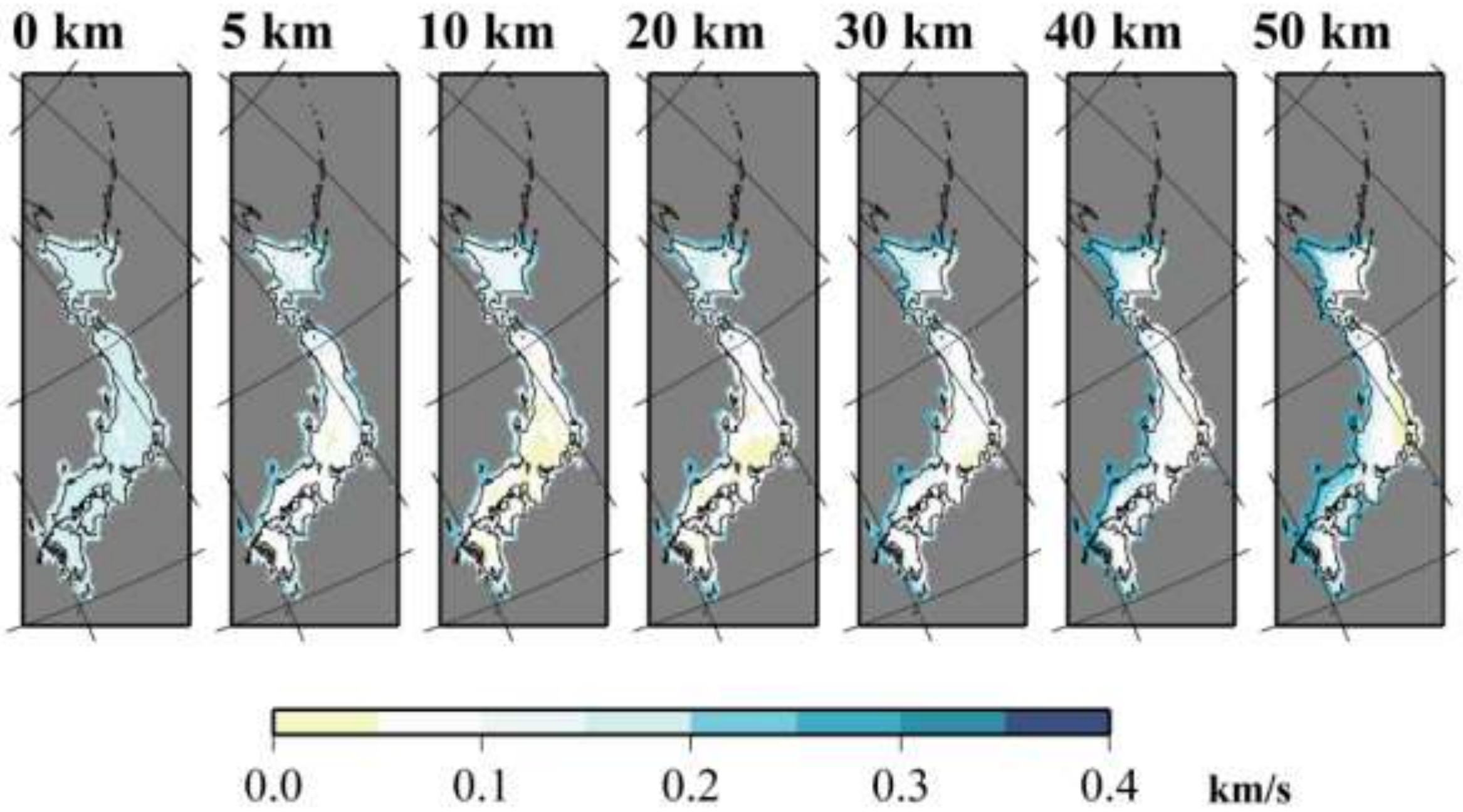

**Figure A1.** Evaluated errors of the tomography model, $\sigma^{obs}$, at 0, 5, 10, 20, 30, 40 and 50 km depth.

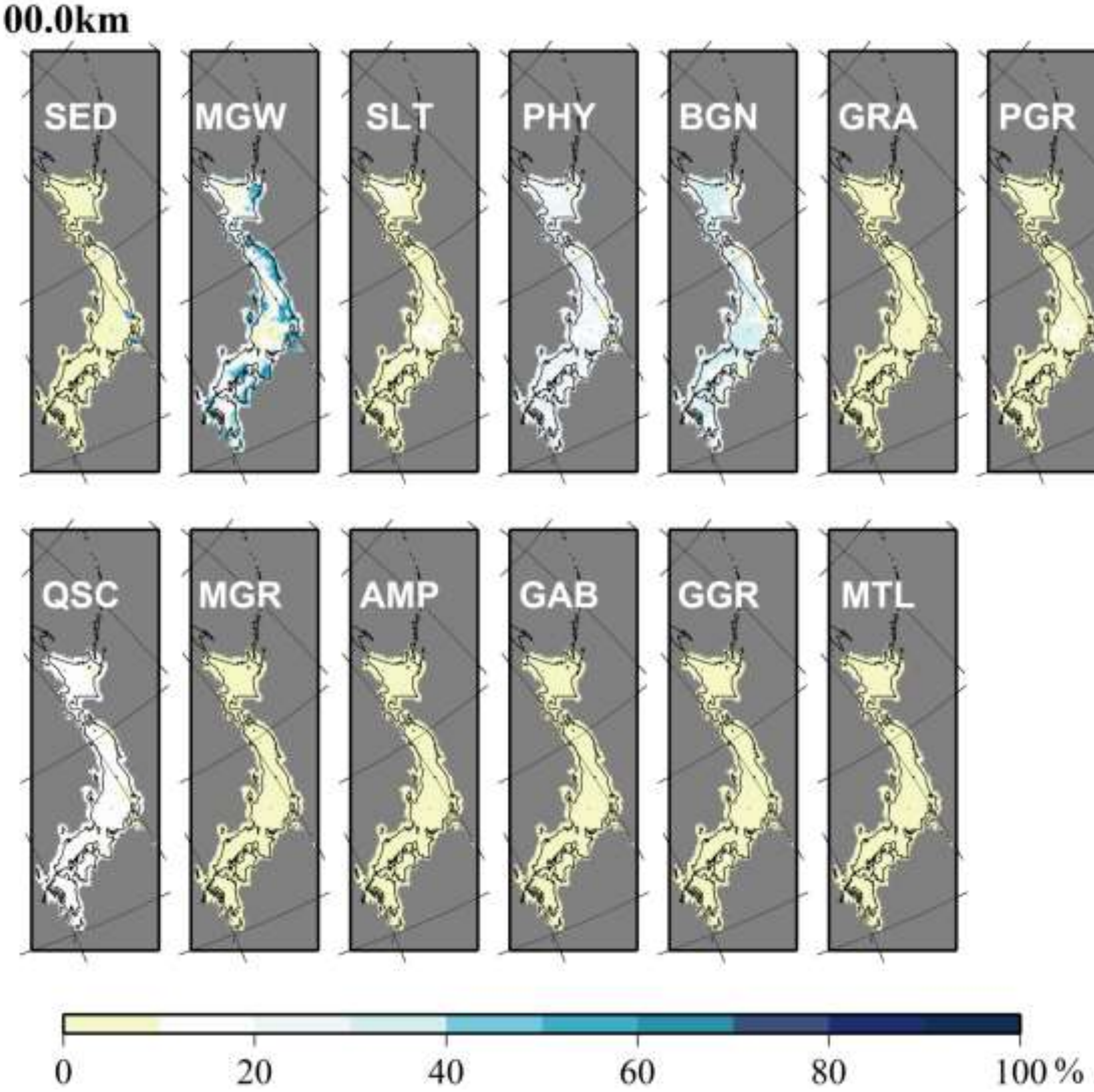

**Figure A2.** The posterior probability $P^{(x)}(i|v^{obs})$ at 0 km depth.

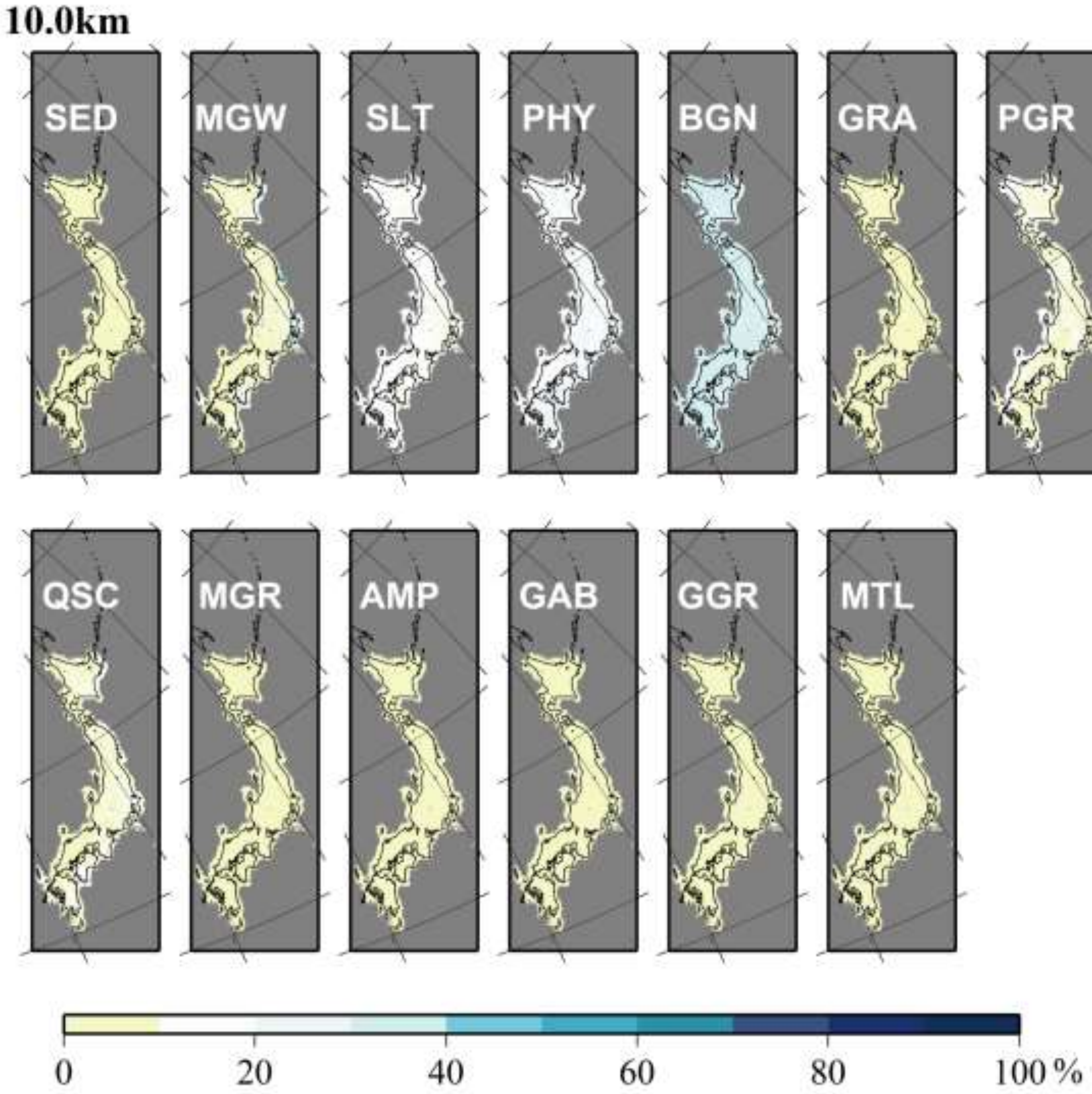

**Figure A2 (continued).** The posterior probability $P^{(x)}(i|v^{obs})$ at 10 km depth.

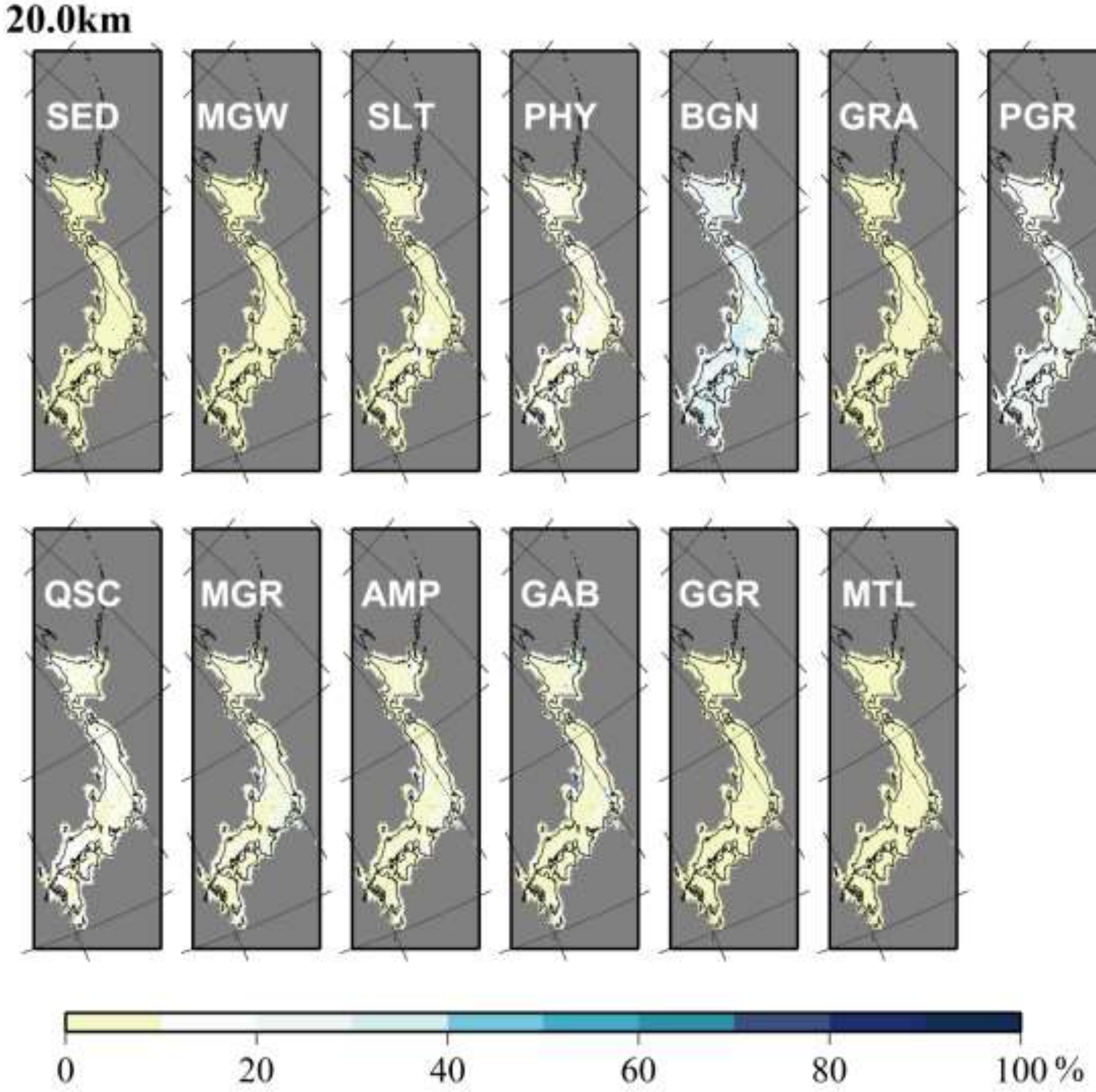

**Figure A2 (continued).** The posterior probability $P^{(x)}(i|v^{obs})$ at 20 km depth.

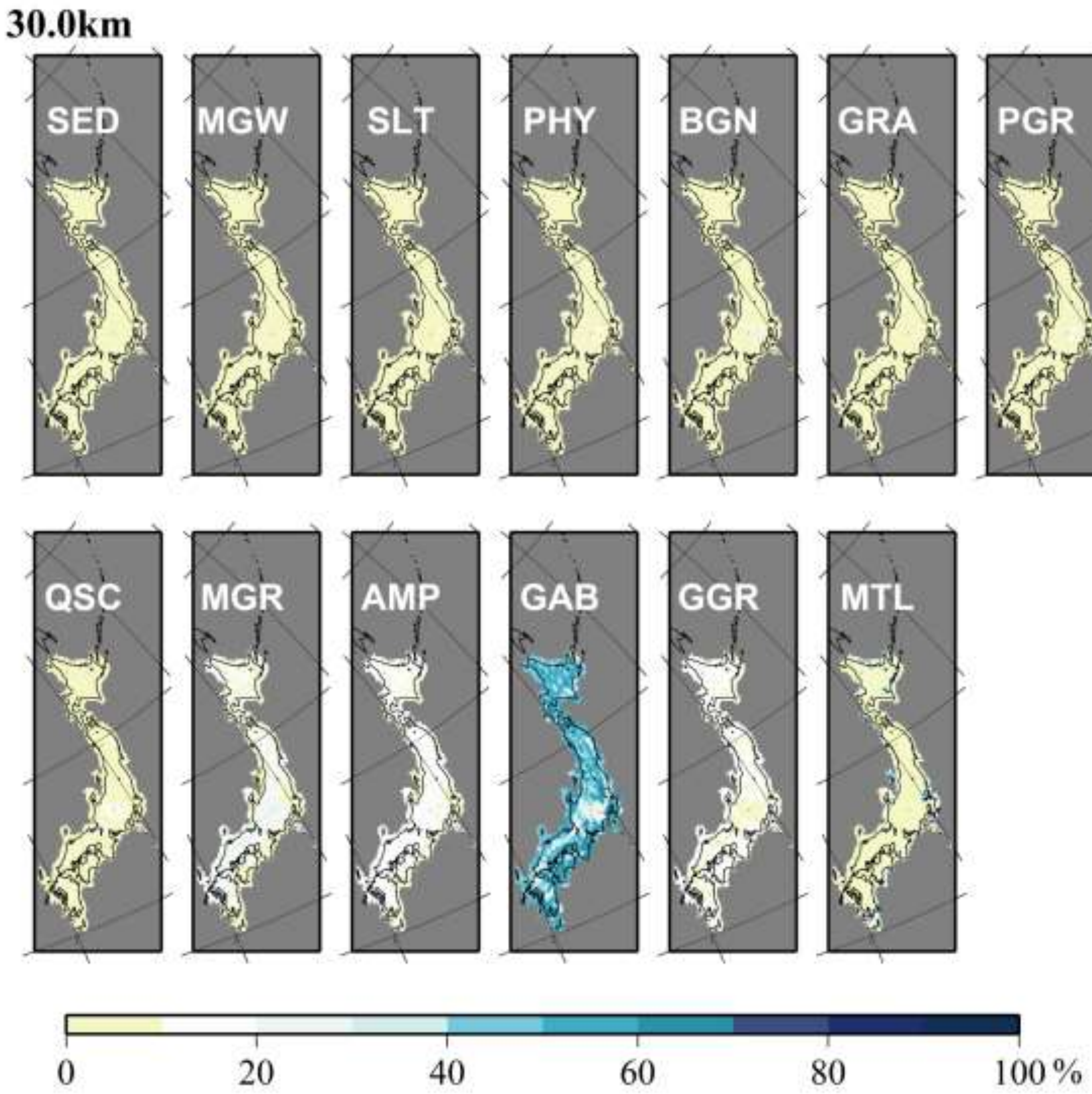

**Figure A2 (continued).** The posterior probability $P^{(x)}(i|v^{obs})$ at 30 km depth.

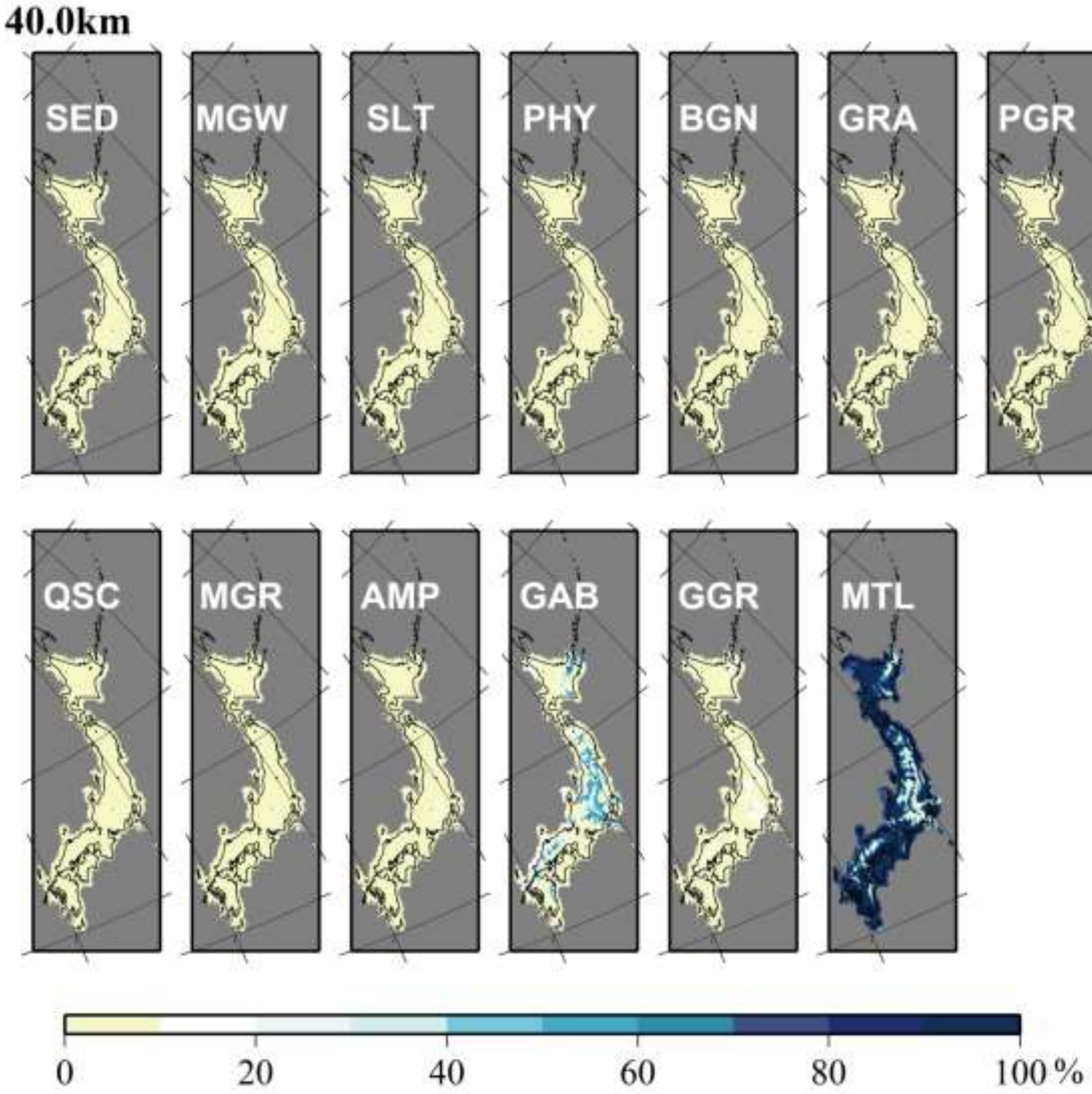

**Figure A2 (continued).** The posterior probability $P^{(x)}(i|v^{obs})$ at 40 km depth.

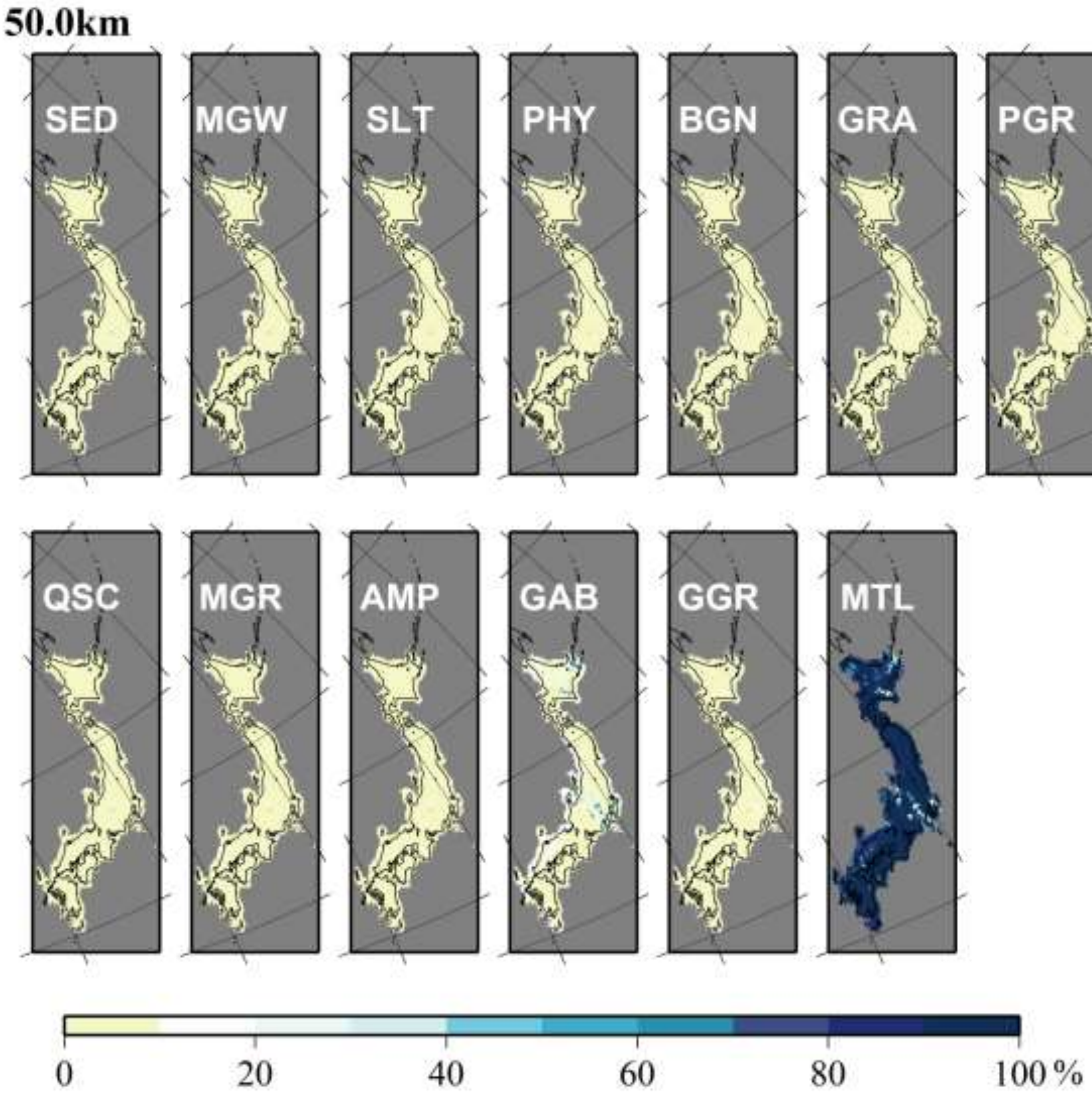

**Figure A2 (continued).** The posterior probability $P^{(x)}(i|v^{obs})$ at 50 km depth.

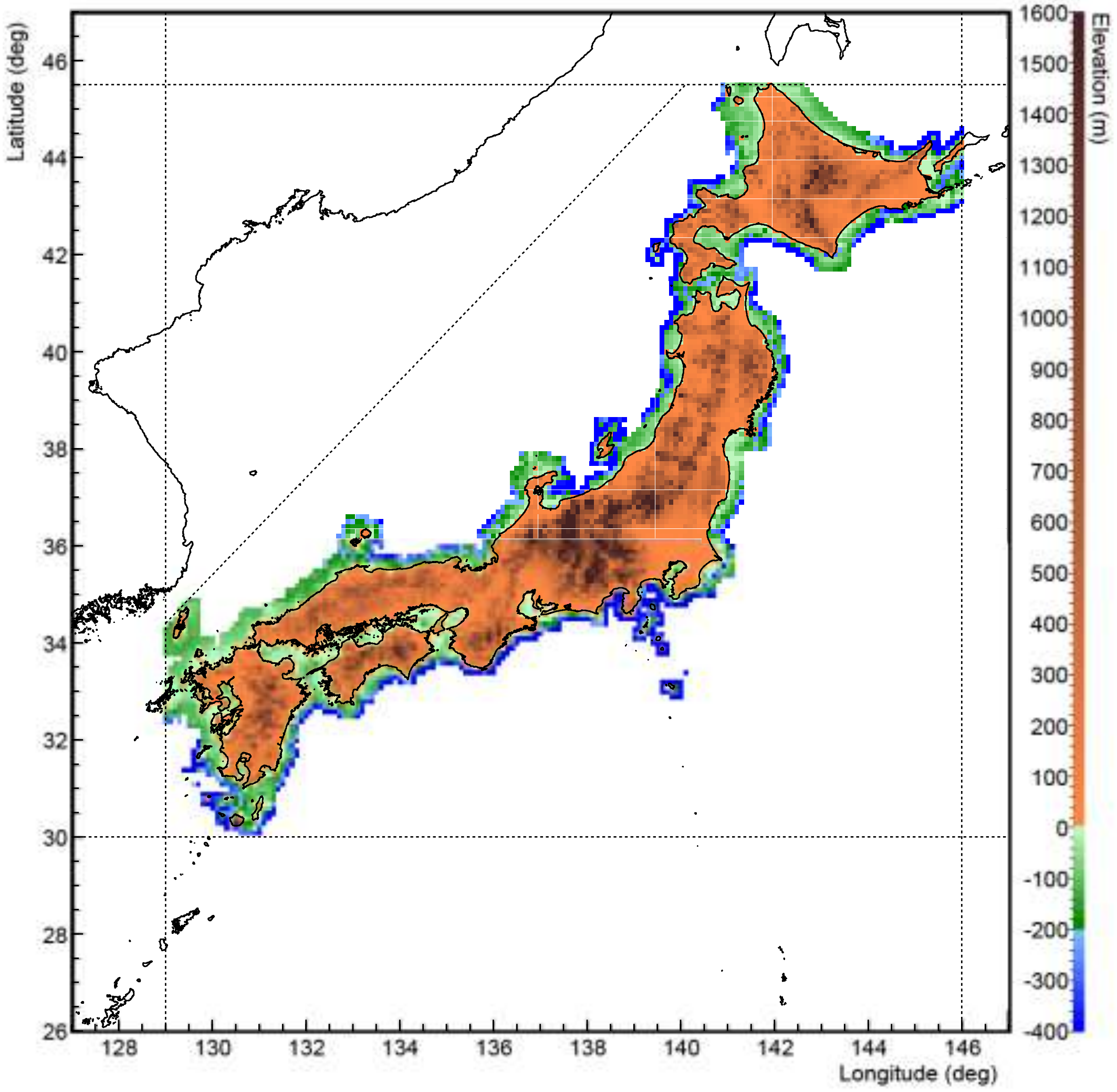

**Figure A3.** The geographical region of interest for this work (shown as colored pixels).

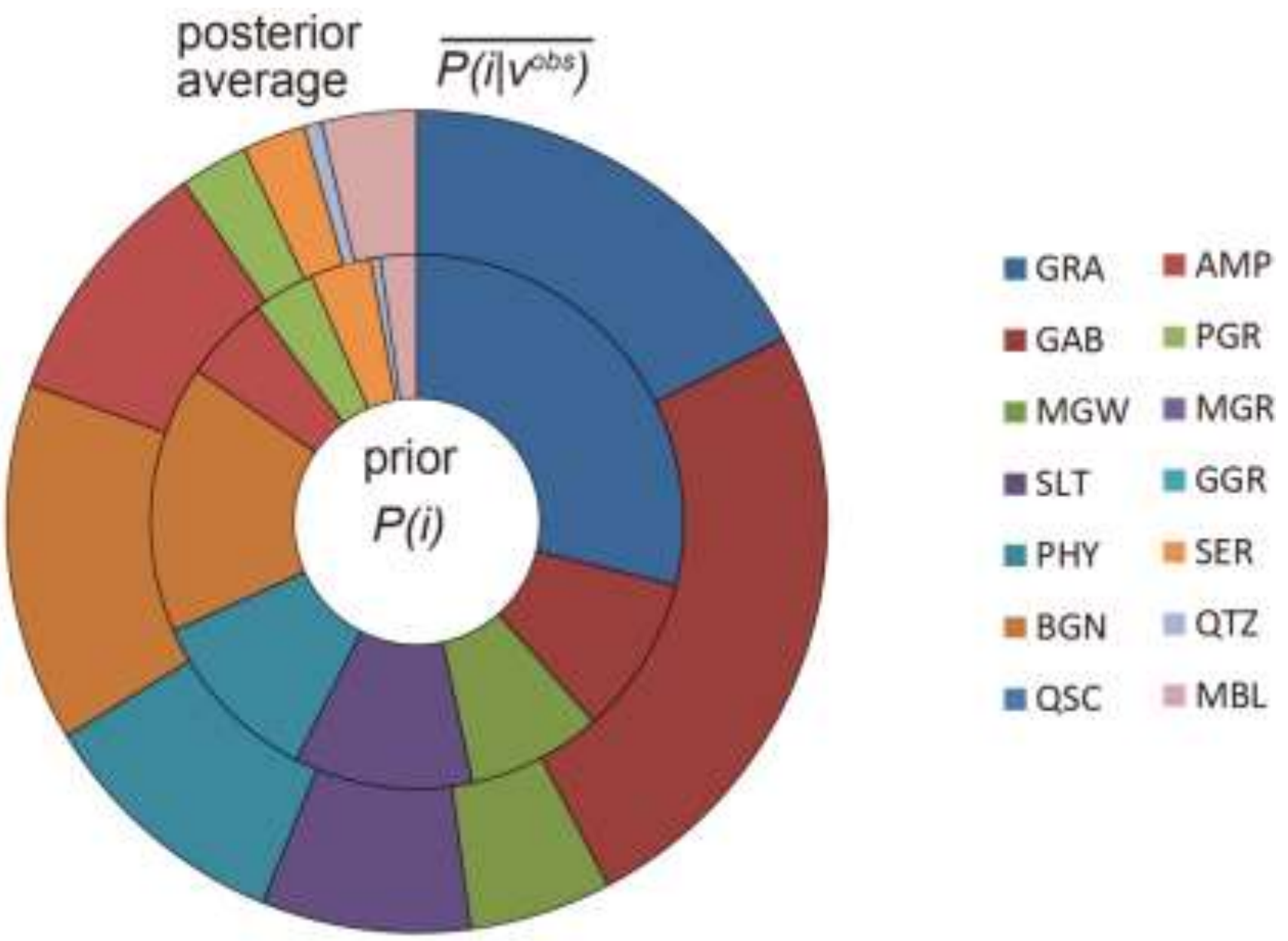

**Figure A4.** Same as Fig. 5 except for using a non-local prior model constructed from the proportions of the North Cascades section (Miller and Paterson, 2001). Comparison with Fig. 5 indicates that the resemblance between the prior and posterior models in Fig. 5 is not a general feature and posterior models can be largely modified from the prior models by the seismological inference. For the case of the North Cascade models, the posterior model is clearly distinct from its prior model, but closer to the mid-point between the North Cascade prior and the Hidaka prior/posterior models, implying that the seismological likelihood is pulling the North Cascade prior model towards the Hidaka model. More quantitatively, when three rocks that appear only in the North Cascade model (SER, QTZ and MBL) are excluded, the Kullback–Leibler (KL) divergence from the Hidaka prior model to the North Cascade prior/posterior models is reduced from 0.55 of the prior model to 0.35 of the posterior model. The KL divergence to the Hidaka posterior model from the Hidaka prior model is 0.02.

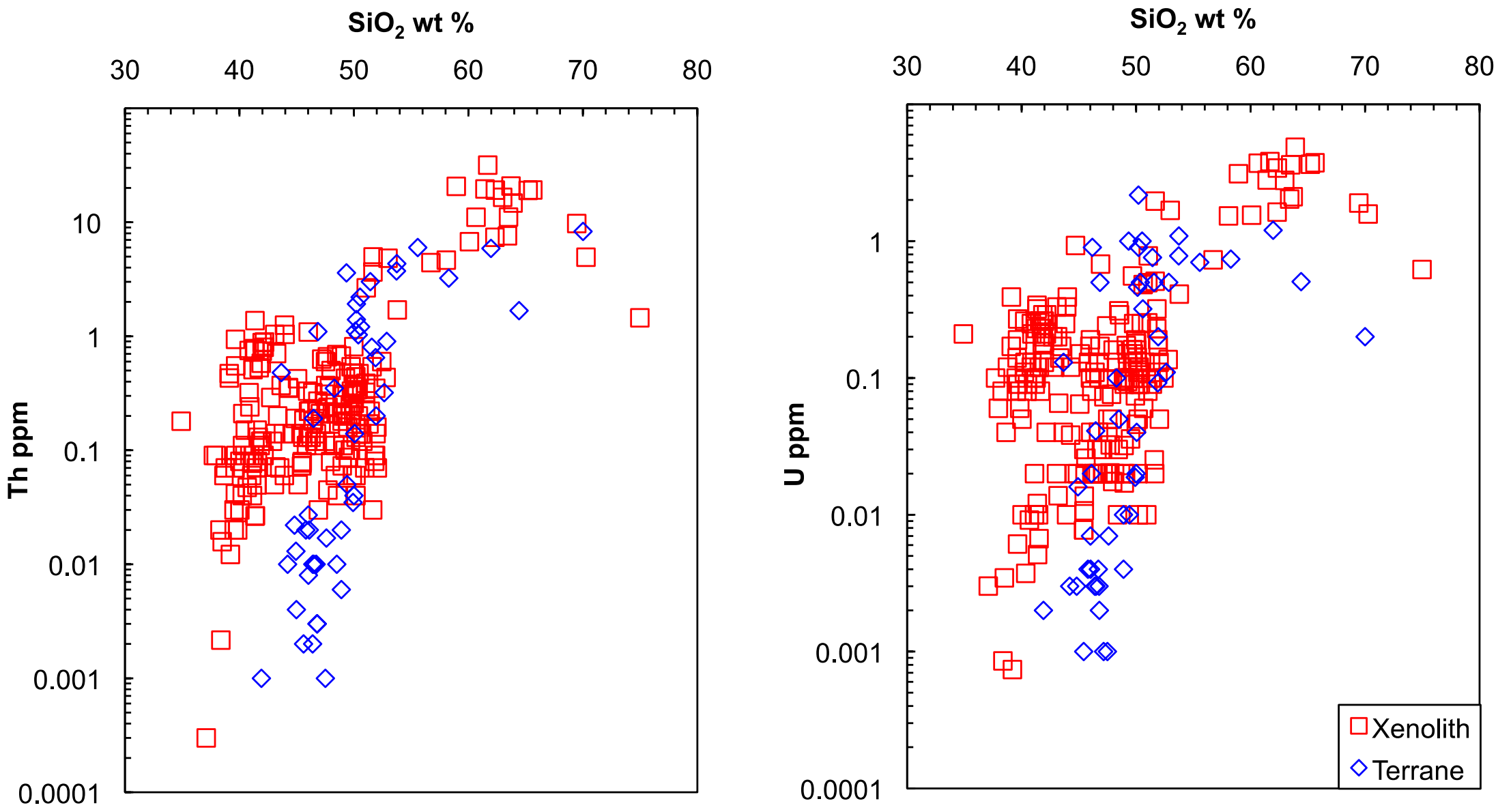

**Figure A5.** U, Th and SiO2 concentrations of granulite and gabbro.

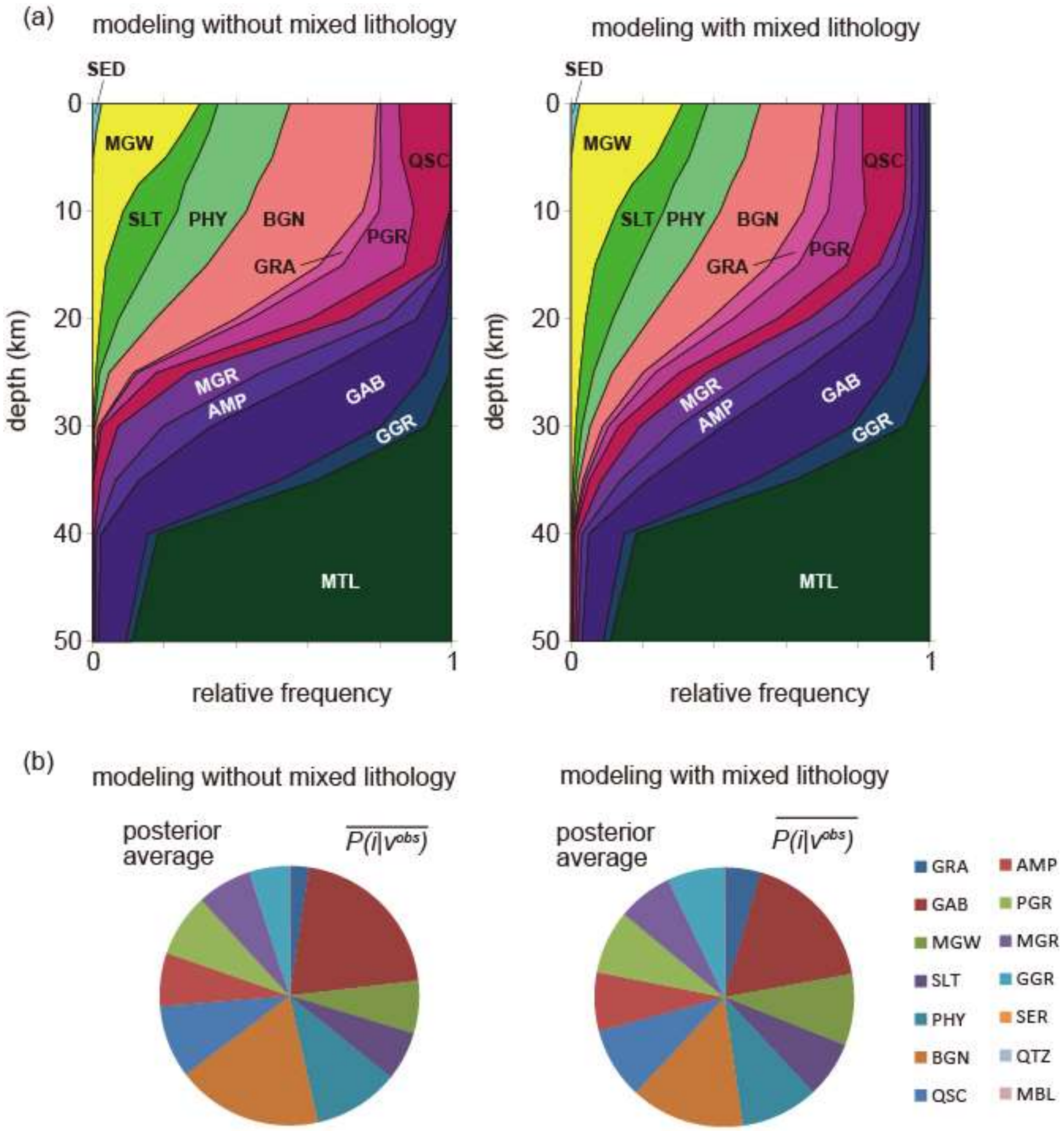

**Figure A6.** (a) Comparison of relative lithology frequencies at each depth between modeling without and with mixed lithology. The left panel is the same as Fig. 3a. (b) Comparison of relative lithology frequencies in the posterior ensemble. The left panel is the same as the right panel in Fig. 5.

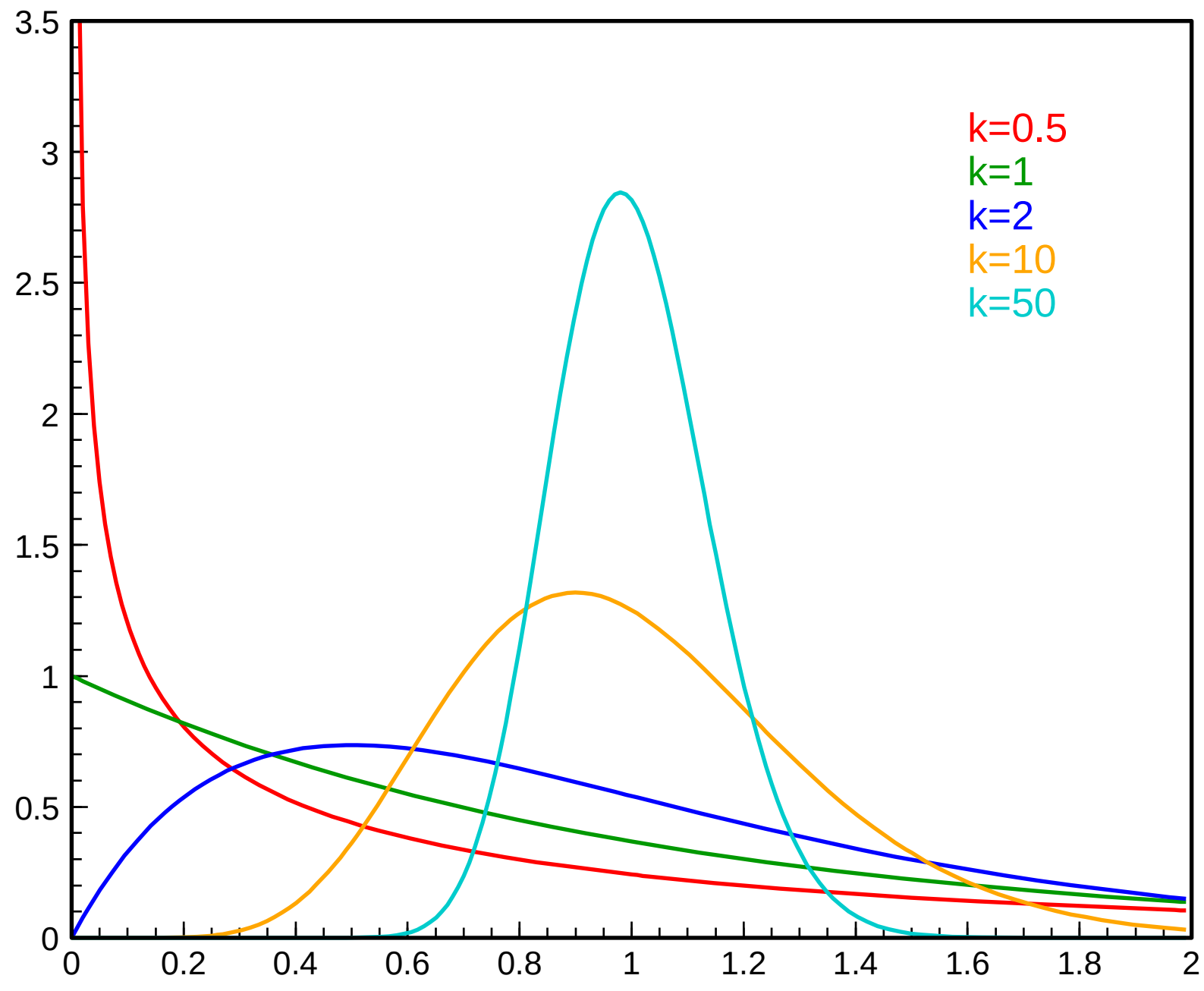

**Figure A7.** Gamma distributions for various shape parameters, for a constant mean value (mean $\equiv k\theta = 1$). If two independent sample pieces with identical concentration distribution, $\mathrm{Gamma}(k, \theta)$, are mixed, the concentration of the merged sample follows $\mathrm{Gamma}(2k, \theta/2)$.

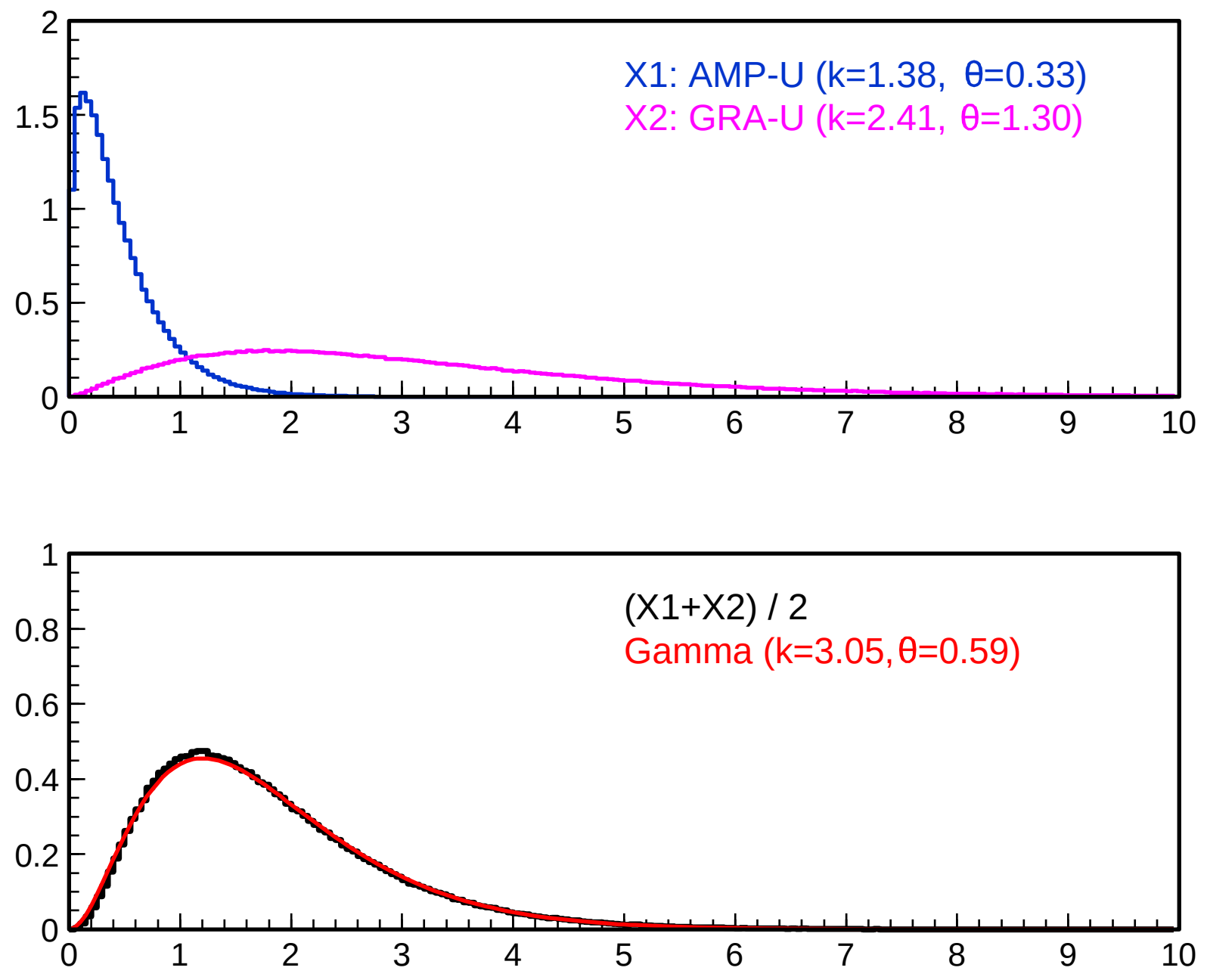

**Figure A8.** Sum of two gamma variates and approximated gamma PDF. In the upper panel, two distinct gamma distributions are shown, one taken from the model of uranium concentration in amphibolite (AMP), the other from uranium in granite-granodiorite (GRA). The lines are normalized distributions of 1,000,000 MC generated samples, X1 and X2, respectively, according to the composition gamma models. In the lower panel, the normalized distribution of the half-sum of the variates, (X1+X2)/2, is shown (black line) together with an approximated gamma distribution calculated with the gamma PDF formula with $k$ and $\theta$ parameters shown in the panel (red line). (X1+X2)/2 corresponds to the uranium concentration in a 1:1 mixing of AMP and GRA. Note that different ratio mixing only changes the shape parameters $k$ of the input distributions as described in the piece-size dependency discussion (Section 6), and the mixing is still a sum of two gamma variates.

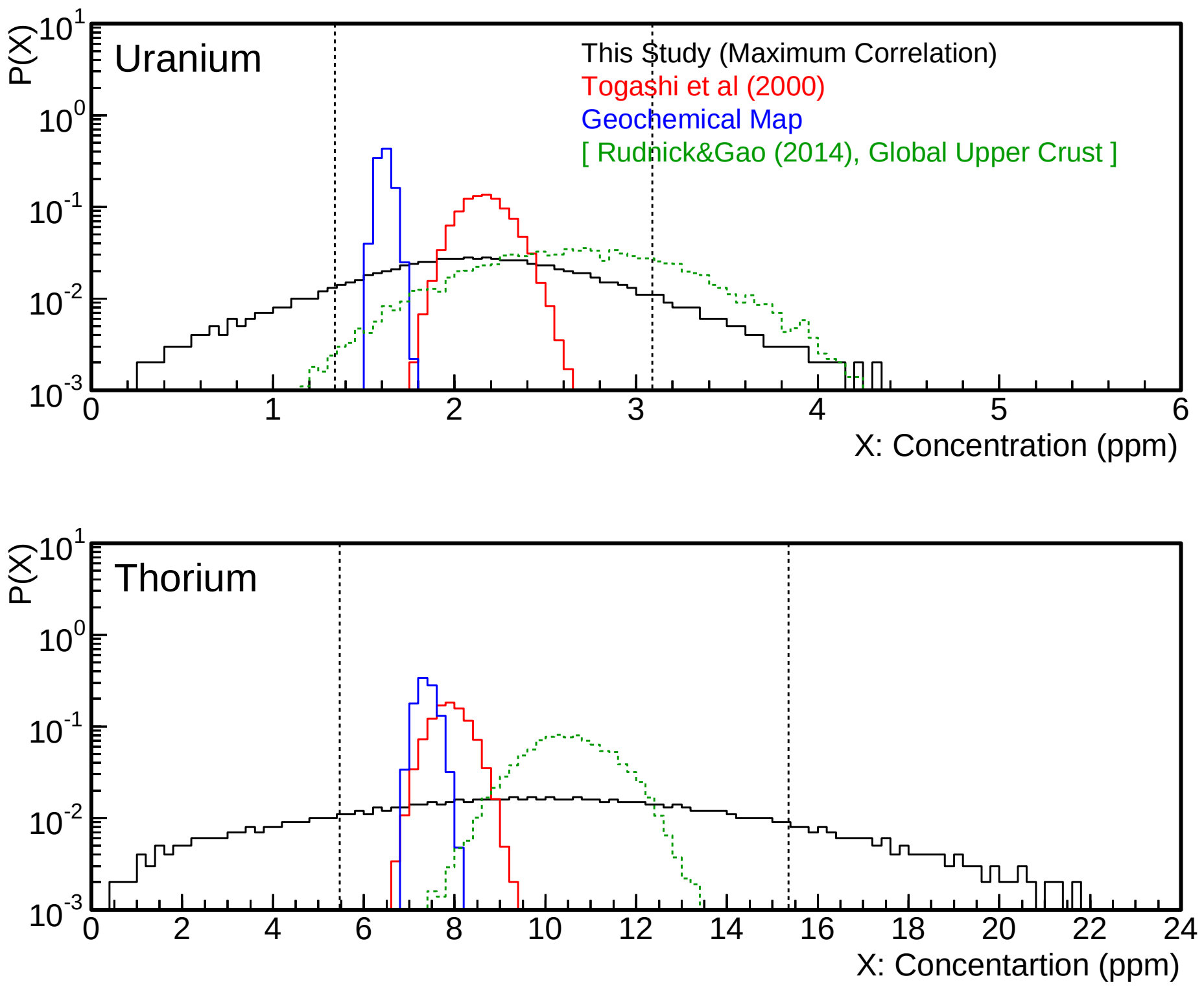

**Figure A9**. Comparison of average surface composition estimations. Upper panel shows the probability density function (PDF) of the average Uranium concentration of Japan arc surface, and lower panel for Thorium. See Appendix A6 for construction of the PDFs. The vertical dotted lines shows the intervals of 68.3% coverage for our model. Log scale on vertical axis is used to make all the distributions visible.